\documentclass[pre,twocolumn,showpacs,superscriptaddress,preprintnumbers,floatfix]{revtex4-1}

\usepackage{etex}
\usepackage{ifpdf}
\usepackage{hyperref}
\usepackage{dcolumn}
\usepackage{url}
\usepackage{amsmath}
\usepackage{amscd}
\usepackage{amsfonts}
\usepackage{amssymb}
\usepackage{amsthm}
\usepackage{xfrac}
\usepackage{braket}
\usepackage{bm}   
\usepackage{bbm}
\usepackage{verbatim}
\usepackage{stmaryrd}
\usepackage{xcolor}
\usepackage{setspace}

\newcommand{\eM}     {\mbox{$\epsilon$-machine}}
\newcommand{\eMs}    {\mbox{$\epsilon$-machines}}
\newcommand{\EM}     {\mbox{$\epsilon$-Machine}}

\newcommand{\Process}{\mathcal{P}}

\newcommand{\MeasAlphabet}  {\mathcal{A}}
\newcommand{\MeasSymbol}   { {X} }
\newcommand{\meassymbol}   { {x} }

\newcommand{\CausalState}   { \mathcal{S} }

\newcommand{\causalstate}   { \sigma }
\newcommand{\CausalStateSet}    { \boldsymbol{\CausalState} }
\newcommand{\AlternateState}    { \mathcal{R} }

\newcommand{\AlternateStateSet} { \boldsymbol{\AlternateState} }

\newcommand{\Prob}      {\Pr} 

\newcommand{\Cmu}       {C_\mu}
\newcommand{\hmu}       {h_\mu}
\newcommand{\EE}        {{\bf E}}
\newcommand{\TI}        {{\bf T}}
\newcommand{\SI}        {{\bf S}}

\newcommand{\PC}        {\chi}

\newcommand{\ProcessAlphabet}   {\MeasAlphabet}

\newcommand{\forward}{+}
\newcommand{\reverse}{-}
\newcommand{\forwardreverse}{\pm} 

\newcommand{\FutureCausalState} { {\CausalState}^{\forward} }

\newcommand{\PastCausalState}   { {\CausalState}^{\reverse} }

\newcommand{\one}{\mathbf{1}}

\newcommand{\lastindex}[2]{
  \edef\tempa{0}
  \edef\tempb{#2}
  \ifx\tempa\tempb
    \edef\tempc{#1}
  \else
    \edef\tempa{0}
    \edef\tempb{#1}
    \ifx\tempa\tempb
      \edef\tempc{#2}
    \else
      \edef\tempc{#1+#2}
    \fi
  \fi
  \tempc
}

\newcommand{\CSjoint}[1][,]{
   \edef\tempa{:}
   \edef\tempb{#1}
   \ifx\tempa\tempb
      \ensuremath{\FutureCausalState\!#1\PastCausalState}
   \else
      \ensuremath{\FutureCausalState#1\PastCausalState}
   \fi
}

\newif\ifpm
\edef\tempa{\forwardreverse}
\edef\tempb{\pm}
\ifx\tempa\tempb
   \pmfalse
\else
   \pmtrue
\fi

\def\clap#1{\hbox to 0pt{\hss#1\hss}}

\begin{document}

\title{Exact Complexity:\\
The Spectral Decomposition of Intrinsic Computation}

\author{James P. Crutchfield}
\email{chaos@ucdavis.edu}
\affiliation{Complexity Sciences Center and Department of Physics, University of
  California at Davis, One Shields Avenue, Davis, CA 95616}
\affiliation{Santa Fe Institute, 1399 Hyde Park Road, Santa Fe, NM 87501}

\author{Christopher J. Ellison}
\email{cellison@wisc.edu}
\affiliation{Center for Complexity and Collective Computation, University of
Wisconsin-Madison, Madison, WI 53706}

\author{Paul M. Riechers}
\email{pmriechers@ucdavis.edu}
\affiliation{Complexity Sciences Center and Department of Physics, University of
  California at Davis, One Shields Avenue, Davis, CA 95616}

\date{\today}
\bibliographystyle{unsrt}

\begin{abstract}
We give exact formulae for a wide family of complexity measures that
capture the organization of hidden nonlinear processes. The spectral
decomposition of operator-valued functions leads to closed-form expressions
involving the full eigenvalue spectrum of the mixed-state presentation of a
process's \eM\ causal-state dynamic. Measures include correlation functions,
power spectra, past-future mutual information, transient and synchronization
informations, and many others. As a result, a direct and complete analysis of
intrinsic computation is now available for the temporal organization of
finitary hidden Markov models and nonlinear dynamical systems with generating
partitions and for the spatial organization in one-dimensional systems,
including spin systems, cellular automata, and complex materials via chaotic
crystallography.

\vspace{0.2in}

\end{abstract}

\preprint{Santa Fe Institute Working Paper 13-XX-XXX}
\preprint{arxiv.org:13XX.XXXX [physics.gen-ph]}

\maketitle


\setstretch{1.1}

\newcommand{\Abet}{\ProcessAlphabet}
\newcommand{\MS}{\MeasSymbol}
\newcommand{\ms}{\meassymbol}
\newcommand{\SSet}{\CausalStateSet}
\newcommand{\St}{\CausalState}
\newcommand{\st}{\causalstate}
\newcommand{\MxSt}{\AlternateState}
\newcommand{\MxSSet}{\AlternateStateSet}
\newcommand{\mxst}{\eta}
\newcommand{\mxstt}[1]{\eta_{#1}}
\newcommand{\StartMS}{\bra{\delta_\pi}}


The emergence of organization in physical, engineered, and social systems is a
fascinating and now, after half a century of active research, widely appreciated phenomenon
\cite{Doob01a,Cros93a,Cama03a,Newm06a,Helb12a}. Success in extending
the long list of instances of emergent organization, however, is not
equivalent to understanding what organization itself is. How do we say
objectively that new organization has appeared? How do we measure
quantitatively how organized a system has become?

\emph{Computational mechanics'} answer to these questions is that a system's
organization is captured in how it stores and processes information---how it
computes \cite{Crut12a}. \emph{Intrinsic computation} was introduced two
decades ago to analyze the inherent information processing in complex systems
\cite{Crut88a}: How much history does a system remember? In what architecture
is that information stored? And, how does the system use it to generate future
behavior?

Computational mechanics, though, is part of a long historical trajectory focused
on developing a physics of information \cite{Weav48a,Bril56a,Land61a}. That
nonlinear systems actively process information goes back to Kolmogorov
\cite{Kolm59}, who adapted Shannon's communication theory \cite{Shan48a}
to measure the information production rate of chaotic dynamical systems. In
this spirit, today computational mechanics is routinely used to determine
physical and intrinsic computational properties in single-molecule dynamics
\cite{Li08a}, in complex materials \cite{Varn12a}, and even in the formation of
social structure \cite{Darm13a}, to mention several recent examples.

Thus, measures of complexity are important to quantifying how organized
nonlinear systems are: their randomness and their structure. Moreover, we now
know that randomness and structure are intimately intertwined. One cannot be
properly defined or even practically measured without the other
\cite[and references therein]{Crut01a}.

Measuring complexity has been a challenge: Until recently, in understanding
the varieties of organization to be captured; still practically, in terms
of estimating metrics from experimental data. One major reason for these
challenges is that systems with emergent properties are hidden: We do not
have direct access to their internal, often high-dimensional state space; we
do not know a priori what the emergent patterns are. Thus, we must
``reconstruct'' their state space and dynamics
\cite{Pack80,Take81,Crut87a,Farm87}. Even then, when successful,
reconstruction does not lead easily or directly to measures of structural
complexity and intrinsic computation \cite{Crut88a}. It gives access to what is
hidden, but does not say what the mechanisms are nor how they work.

Our view of the various kinds of complexity and their measures, though, has
become markedly clearer of late. There is a natural semantics of complexity in which
each measure answers a specific question about a system's organization.
For example:
\begin{itemize}
\setlength{\itemsep}{-4pt}
\setlength{\topsep}{-6pt}
\setlength{\parsep}{-6pt}
\setlength\itemindent{-15pt}
\item How random is a process?
	Its \emph{entropy rate} $\hmu$ \cite{Kolm59}.
\item How much information is stored?
	Its \emph{statistical complexity} $\Cmu$ \cite{Crut88a}.
\item How much of the future can be predicted?
	Its \emph{past-future mutual information} or \emph{excess entropy} $\EE$
	\cite{Crut01a}.
\item How much information must an observer extract to know a process's hidden
	states? Its \emph{transient} information $\TI$ and
	\emph{synchronization information} $\SI$ \cite{Crut01a}.
\item How much of the generated information ($\hmu$) affects
	future behavior? Its \emph{bound information} $b_\mu$ \cite{Jame11a}.
\item What's forgotten? Its \emph{ephemeral information} $\rho_\mu$
	\cite{Jame11a}.
\end{itemize}
And there are other useful measures ranging from degrees of irreversibility
to quantifying model redundancy; see, for example,
Ref. \cite{Crut08a}
and the proceedings in Refs. \cite{Marc11a,Crut11a}.

Unfortunately, except in the simplest cases where expressions are known for
several, to date typically measures of intrinsic computation require extensive
numerical simulation and estimation. Here we answer this challenge, providing
exact expressions for a process's measures in terms of its \eM. In particular,
we show that the spectral decomposition of this hidden dynamic leads to
closed-form expressions for complexity measures. In this way, analyzing
intrinsic computation reduces to mathematically constructing or reliably
estimating a system's \eM.


Our main object of study is a process $\Process$, by which we mean the rather
prosaic listing of all of a system's behaviors or realizations
$\{ \ldots \ms_{-2}, \ms_{-1}, \ms_{0}, \ms_{1}, \ldots \}$ and their
probabilities: $\Prob(\ldots \MS_{-2}, \MS_{-1}, \MS_{0}, \MS_{1}, \ldots)$.
We assume the process is stationary and ergodic and the measurement values
range over a finite alphabet: $\ms \in \Abet$. This class describes a wide
range of processes from statistical mechanical systems in equilibrium and in
nonequilibrium steady states to nonlinear dynamical systems in discrete and
continuous time on their attracting invariant sets.

Following Shannon and Kolmogorov, information theory gives a
natural measure of a process's randomness as the uncertainty in measurement
blocks: $H(L) = H \left[ \MS_{0:L} \right]$, where $H$ is the
Shannon-Boltzmann entropy of the distribution governing the block
$\MS_{0:L} = \MS_0, \MS_1, \ldots, \MS_{L-1}$. We monitor the
\emph{block entropy} growth using:
\begin{align}
\hmu (L) & = H(L) - H(L-1) \nonumber \\
  & = H[ \MS_{L-1} | \MS_{0:L-1} ] ~,
\label{eq:EntropyRateEst}
\end{align}
where the latter is the uncertainty in the next measurement $\MS_{L-1}$
conditioned on knowing the preceding block $\MS_{0:L-1}$. And when the limit
exists, we say the process generates information at the \emph{entropy rate}:
$\hmu = \lim_{L \to \infty} \hmu (L)$.

Measurements, though, only indirectly reflect a system's internal
organization. Computational mechanics extracts that hidden organization via
the process's \eM\ \cite{Crut12a},
consisting of a set of recurrent
\emph{causal states} $\SSet = \{\st^0, \st^1, \st^2, \ldots\}$ and transition
dynamic $\{ T^{(\ms)}: \ms \in \Abet \}$. The \eM\ is a system's unique,
minimal-size, optimal predictor from which two key complexity measures can
be directly calculated.

The entropy rate follows immediately from the \eM\ as the causal-state
averaged transition uncertainty:
\begin{align}
\hmu = - \sum_{\st \in \SSet} \Prob(\st)
  \sum_{\ms \in \Abet} \Prob(\ms|\st) \log_2 \Prob(\ms|\st) ~.
\label{eq:UnifilarEntropyRate}
\end{align}
Here, the causal state distribution $\Prob(\St)$ is the stationary distribution
$\bra{\pi} = \bra{\pi} T$ of the internal Markov chain governed by the
row-stochastic matrix $T = \sum_{\ms \in \Abet} T^{(\ms)}$. The conditional
probabilities $\Prob(\ms|\st)$ are the associated transition components in
the labeled matrices $T_{\st, \st^\prime}^{(\ms)}$. Note that the next state
$\st^\prime$ is uniquely determined by knowing the current state $\st$ and
the measurement value $\ms$---a key property called \emph{unifilarity}.

The amount of historical information the process stores also follows
immediately: the \emph{statistical complexity}, the Shannon-Boltzmann entropy
of the causal-state distribution:
\begin{align}
\Cmu = - \sum_{\st \in \SSet} \Prob(\st) \log_2 \Prob(\st) ~,
\label{eq:StatisticalComplexity}
\end{align}

In this way, the \eM\ allows one to directly determine two important properties of a
system's intrinsic computation: its information generation and its storage.
Since it depends only block entropies, however, $\hmu$ can be
calculated via other presentations; though not as efficiently. For example,
$\hmu$ can be determined from Eq.  (\ref{eq:UnifilarEntropyRate}) using any
unifilar predictor, which necessarily is always larger than the \eM.
Only recently was a (rather more complicated) closed-form expression discovered
for the excess entropy $\EE$ using a representation closely related to the
\eM~\cite{Crut08a}. Details aside, no analogous closed-form expressions
for the other complexity measures are known, including and especially those for
finite-$L$ blocks, such as $\hmu(L)$.


To develop these, we shift to consider how an observer represents its knowledge
of a hidden system's current state and then introduce a spectral analysis of
that representation. For our uses here, the observer has a correct model
in the sense that it reproduces $\Process$ exactly. (Any model that does we
call a \emph{presentation} of the process. There may be many.) Using this,
the observer tracks a process's evolution using a distribution over the
hidden states called a \emph{mixed state}
$\mxst \equiv \left( \Prob(\st^0), \Prob(\st^1), \Prob(\st^2), \ldots \right)$.
The associated random variable is denoted $\MxSt$. The question is how does an
observer update its knowledge ($\mxst$) of the internal states as it makes
measurements---$\ms_{0}, \ms_{1}, \ldots$?

If a system is in mixed state $\mxst$, then the probability of seeing
measurement $\ms$ is:
$\Prob(\MS = \ms| \MxSt = \mxst) = \bra{\mxst} T^{(\ms)} \ket{\one}$,
where $\bra{\mxst}$ is the mixed state as a row vector and $\ket{\one}$ is
the column vector of all $1$s. This extends to measurement sequences
$w = \ms_0 \ms_1 \ldots \ms_{L-1}$, so that if, for example, the process is
in statistical equilibrium,
$\Prob(w) = \bra{\pi} T^{(w)} \ket{\one} = \bra{\pi} T^{(\ms_0)}
T^{(\ms_1)} \dotsm T^{(\ms_{L-1})} \ket{\one}$.
The mixed-state evolution induced by measurement sequence $w$ is:
$\bra{\mxstt{t+L}}
  = \bra{\mxstt{t}} T^{(w)} / \bra{\mxstt{t}} T^{(w)} \ket{\one}$.
The set $\MxSSet$ of mixed states that we use here are those induced by all
allowed words $w \in \Abet^*$ from initial mixed state $\mxstt{0} = \pi$.
For each mixed state $\mxstt{t+1}$ induced by symbol $x \in \Abet$,
the mixed-state-to-state transition probability is:
$\Prob \left( \mxstt{t+1}, \ms \middle| \mxstt{t} \right)
  = \Prob \left( \ms \middle| \mxstt{t} \right)$.
And so, by construction, using mixed states gives a unifilar presentation.
We denote the associated set of transition matrices $\{W^{(\ms)}\}$. They and
the mixed states $\MxSSet$ define a process's
\emph{mixed-state presentation} (MSP), which describes how an observer's
knowledge of the hidden process updates via measurements. The row-stochastic
matrix $W = \sum_{\ms \in \Abet} W^{(\ms)}$ governs the evolution of the
probability distribution over allowed mixed states.

The use of mixed states is originally due to Blackwell \cite{Blac57b}, who
expressed the entropy rate $\hmu$ as an integral of a (then uncomputable)
measure over the mixed-state state space $\MxSSet$.
Although we focus here on the finite mixed-state case for simplicity, it is
instructive to see in the general case the complicatedness revealed in a
process using the mixed-state presentation: e.g., Figs. 17(a)-(c) of Ref.
\cite{Crut92c}. The
Supplementary Materials give the detailed calculations for the finite case.

Mixed states allow one to derive an efficient expression for the finite-$L$
entropy-rate estimates of Eq.\ (\ref{eq:EntropyRateEst}):
\begin{align}
\hmu(L) &= H\bigl[\MS_{L-1} | \bigl(\MxSt_{L-1} | \MxSt_0 = \pi \bigr) \bigr]
  ~.
\label{eq:MixedStateEntropyRateConvergence}
\end{align}
This says that one need only update the initial distribution over mixed
states (with all probability density on $\mxstt{0} = \pi$)
to the distribution at time $L$ by tracking powers $W^L$ of the internal
transition dynamic of the MSP and not tracking, for
that matter, an exponentially growing number of intervening sequences
$\{\ms^L\}$. (This depends critically on the MSP's unifilarity.)
That is, using the MSP reduces the original exponential computational
complexity of estimating the entropy rate to polynomial time in $L$. Finally,
and more to the present task, the mixed-state simplification is the main lead
to an exact, closed-form analysis of complexity measures, achieved by
combining the MSP with a spectral decomposition of the mixed-state evolution
as governed by $W^L$.


State distribution evolution involves iterating the transition dynamic
$W^L$---that is, taking powers of a row-stochastic square matrix. As is well
known, functions of a diagonalizable matrix can often be carried out
efficiently by operating on its eigenvalues. More generally, using the Cauchy
integral formula for operator-valued functions \cite{Hass99a} and given $W$'s
eigenvalues
$\Lambda_W \equiv \{ \lambda \in \mathbb{C}: \text{det} (\lambda I - W) = 0 \}$,
we find that $W^L$'s spectral decomposition is:
\begin{align}
W^L \!  & = \!  \sum_{\lambda \in \Lambda_W \atop \lambda \neq 0}
  \lambda^L W_\lambda
  \left\{
	I + \! \sum_{N=1}^{\nu_\lambda-1} {L \choose N}
	\left( \lambda^{-1} W - I \right)^N
  \right\} \nonumber \\
  & ~~~~+ \left[ 0 \in \Lambda_W \right]
  \left\{
  	\delta_{L,0} W_0 + \! \sum_{N=1}^{\nu_0-1} \delta_{L,N} W_0 W^N
  \right\}
  \! ,
\label{eq:GeneralSpectralDecomp}
\end{align}
where $[0 \in \Lambda_W]$ is the Iverson bracket (unity when $\lambda = 0$
is an eigenvalue, vanishing otherwise), $\delta_{i,j}$ is the Kronecker delta
function, and $\nu_\lambda$ is the size of the largest Jordan block associated
with $\lambda$: $\nu_\lambda \leq 1 + a_\lambda - g_\lambda$,
where $g_\lambda$ and $a_\lambda$ are $\lambda$'s geometric (subspace
dimension) and algebraic (order in the characteristic polynomial)
multiplicities, respectively. The matrices $\{W_\lambda\}$ are a mutually
orthogonal set of projection operators given by the residues of $W$'s resolvent:
\begin{align}
\label{eq:ProjOperatorsViaResiduesOfResolvent}
W_\lambda & =
  \tfrac{1}{2\pi i} \oint_{C_\lambda} (z I - W)^{-1} dz
  ~,
\end{align}
a counterclockwise integral around singular point $\lambda$.

For simplicity here, consider only $W$s that are diagonalizable. In this case:
$g_\lambda = a_\lambda$ and Eq.\ (\ref{eq:GeneralSpectralDecomp}) simplifies to
$W^L = \sum_{\lambda \in \Lambda_W} \lambda^L W_\lambda$,
where the projection operators reduce to
$W_\lambda = \prod_{\zeta \in \Lambda_W \atop \zeta \neq \lambda}
  \left(W - \zeta I\right)/\left(\lambda - \zeta\right)$.
Thus, the only $L$-dependent operation in forming $W^L$ is simply
exponentiating its eigenvalues. The powers determine all of a process's
properties, both transient (finite-$L$) and asymptotic.


Forming the mixed-state presentation of process's \eM, its spectral
decomposition leads directly to analytic, closed-form expressions for many
complexity measures---here we present formulae only for $\hmu(L)$,
$\EE$, $\SI$, and $\TI$. Similar expressions for correlation functions and
power spectra, partition functions, $b_\mu$, $r_\mu$,
and others are presented elsewhere.

Starting from its mixed-state expression in
Eq.\ (\ref{eq:MixedStateEntropyRateConvergence}) for the length-$L$
entropy-rate approximates $h_\mu(L)$, we find the closed-form expression:
\begin{align}
\hmu(L)
  & = \StartMS W^{L-1} \ket{H(W^\Abet) } \nonumber \\
  & = \sum_{\lambda \in \Lambda_W} \lambda^{L-1}
  \StartMS W_\lambda \ket{H(W^\Abet) } ~,
\label{eq:SpectralEntropyRateConvergence}
\end{align}
where $\delta_\pi$ is the
distribution with all probability density on the MSP's unique start state---the
mixed state corresponding to the \eM's equilibrium distribution $\pi$. In
addition, $\ket{H(W^\Abet) }$ is a column vector of transition uncertainties
from each allowed mixed state $\eta$:
\begin{align*}
\ket{H(W^\Abet) } =
  - \!\!\!
  \sum_{\eta \in \MxSSet}
  \ket{\delta_\eta}
  \sum_{x \in \Abet}
  \bra{\delta_\eta} W^{(x)} \ket{\one}
  \log_2
  \bra{\delta_\eta} W^{(x)} \ket{\one} .
\end{align*}
Taking $L \to \infty$, one finds the entropy rate (cf. Eq.
(\ref{eq:UnifilarEntropyRate})):
\begin{align*}
\hmu = \bra{\delta_\pi} W_1 \ket{H(W^\Abet)}
     = \braket{\pi_W | H(W^\Abet)}
	 ~,
\end{align*}
where $\pi_W$ is the stationary distribution over the MSP.

Let's turn to analyze the past-future mutual information, the excess entropy
$\EE = I[\MS_{-\infty:0};\MS_{0:\infty}]$: the information from the
past that reduces uncertainty in the future. In general, $\EE$ is not the
statistical complexity $\Cmu$, which is the information from the past
that must be stored in order make optimal predictions about the future.
Although Eq.\ (\ref{eq:StatisticalComplexity}) makes it clear
that the stored information $\Cmu$ is immediately calculable from the \eM,
$\EE$ is substantially less direct. To see this, recall that the excess
entropy has an equivalent
definition---$\EE = \sum_{L = 1}^\infty \left[\hmu(L) - \hmu \right]$---to
which we can apply Eq.\ (\ref{eq:SpectralEntropyRateConvergence}), obtaining:
\begin{align}
\EE =  \sum_{\lambda \in \Lambda_W \atop |\lambda| < 1}
  \frac{1}{1-\lambda} \StartMS W_\lambda | H(W^\Abet)
  \rangle
  ~.
\label{eq:SpectralExcessEntropy}
\end{align}
This should be compared to the only previously known general closed-form,
which uses a process and its time-reversal \cite{Crut08a,Crut08b}:
$\EE = I[\PastCausalState;\FutureCausalState]$,
where $\PastCausalState$ and $\FutureCausalState$ are the causal states of the
reverse-time and forward-time \eMs, respectively. Thus,
Eq.\ (\ref{eq:SpectralExcessEntropy})---the spectral expression from the
forward process---captures aspects of the reverse-time process.

How does an observer come to know a hidden process's internal state? We
monitor this via the \emph{average state uncertainty} having seen all length-$L$
words \cite{Crut01a}:
\begin{align*}
\mathcal{H}(L) & = - \sum_{w \in \Abet^L} \Prob(w)
  \sum_{\st \in \SSet} \Prob(\st|w) \log_2 \Prob(\st|w) \\
  & =
  \sum_{\eta \in \MxSSet}
  \Prob (\MxSt_L = \eta | \MxSt_0 = \pi ) H[\eta] ~,
\end{align*}
where the last line is the mixed-state version with $H[\eta]$
being the presentation-state uncertainty specified by the mixed state $\eta$.
Applying the spectral decomposition yields, for diagonalizable $W$:
\begin{align*}
\mathcal{H}(L) = \bra{\delta_\pi} W^L \ket{H[\eta]}
  = \sum_{\lambda \in \Lambda_W} \lambda^L
	\bra{\delta_\pi} W_\lambda \ket{H[\eta]}
	~,
\end{align*}
where $\ket{H[\eta]}$ is the column vector of state-distribution uncertainties
for each allowed mixed state $\eta \in \MxSSet$.

The observer is \emph{synchronized} when the state uncertainty vanishes:
$\mathcal{H}(L) = 0$. The total amount of state information, then, that an
observer must extract to become synchronized is the
\emph{synchronization information} \cite{Crut01a}
$\SI \equiv \sum_{L=0}^\infty \mathcal{H}(L)$.
Applying the above spectral decomposition results in the following closed-form
expression:
\begin{align}
\label{eq:SyncInfo}
\SI = 
  \sum_{\lambda \in \Lambda_W \atop |\lambda| < 1}
  \frac{1}{1-\lambda} \StartMS W_\lambda | H[\eta]
  \rangle 
	~.
\end{align}

This form makes it clear that mixed states and the transitions between them
capture fundamental properties of the underlying process. For example, rewriting
$\Cmu$ as the entropy of the initial mixed
state---$\braket{\delta_\pi | H[\eta]}$---reinforces this observation. A
related measure, the presentation-dependent synchronization
information~\cite{Crut10a}, diverges when the presentation is not
synchronizable. Note also the close similarity between the excess entropy
formula Eq. (\ref{eq:SpectralExcessEntropy}) above and Eq. \eqref{eq:SyncInfo}.
The only difference is that they average different informational
quantities---the transition or the state uncertainties, respectively.

Although there are a number of additional complexity measures, as we discussed
above, the final example we present is the \emph{transient information} $\TI$
\cite{Crut01a}. It measures the amount of information one must extract from
observations so that the block entropy converges to its linear asymptote:
$\TI \equiv \sum_{L = 1}^\infty L \left[\hmu(L) - \hmu \right]$.
The spectral decomposition readily yields the following closed-form expression:
\begin{align*}
\TI =  \sum_{\lambda \in \Lambda_W \atop |\lambda| < 1}
  \frac{1}{(1-\lambda)^2} \StartMS W_\lambda | H(W^\Abet) \rangle
  ~.
\end{align*}
This form reveals the close relationship between transient information and
excess entropy: they differ only in the eigenvalue weighting.

There are a number of comments to make at this point to draw out the
results' usefulness and import. A process's structural complexity is not
controlled by only the first spectral gap---the difference between the maximum
and next eigenvalue. Rather, the \emph{entire} spectrum is implicated in
calculating the complexity measures. In terms of the temporal dynamics, all
subspaces of the underlying causal-state process contribute. Naturally, there
will be cases in which only some subspaces dominate, but as the expressions
show this is not the general case. In addition, there is much structural
information to be extracted from the projection operators $W_\lambda$, such
as the dimension (geometric multiplicity) of the associated subspaces on
which they act. This, in turn, gives the number of active degrees of freedom
for the constituent subprocesses. As a result, we see that complexity
measures capture inherently different properties---transient, finite-time,
and time asymptotic---far beyond correlations and power spectra.

Although their derivations have not been laid out, the formulae as given
are immediately usable. The Supplementary Materials provide calculations of
complexity measures for several examples, including those for processes
generated by typical unifilar hidden Markov models, nonlinear dynamical systems,
cellular automata, and materials that form close-packed structures, as
determined experimentally from X-ray diffraction spectra.


One of the more direct physical consequences of computational mechanics is
that a system's organization is synonymous with how it stores and transforms
information---a view complementary to that physics takes in terms of energy
storage and transduction. In short, how a system is organized is how it
computes. The theory just introduced grounds this information processing view
practically and mathematically, giving quantitative and exact analysis of how
hidden processes are organized. And, it should be contrasted
with Kolmogorov-Chaitin complexity analysis \cite{Vita93a}. For both
stochastic and deterministic chaotic processes, the Kolmogorov-Chaitin
complexity is (i) dominated by randomness and (ii) uncomputable.
The theory here could not be more different: A wide variety of distinct kinds
of information storage and processing are identified and they are exactly
calculable.


The authors thank Ryan James and Dowman Varn for helpful discussions. This work
was partially supported by ARO grant W911NF-12-1-0234 and by a grant from the
John Templeton Foundation.

\bibliography{chaos}

\newpage

\appendix

\section*{Supplementary Materials}

The following sections provide detailed calculations of the mixed-state
presentations and the measures of intrinsic computation discussed in the main
text using the closed-form expressions there.
We analyze first a prototypic strictly sofic system,
the Even Process, and then give results for the symbolic
dynamics of the Tent Map at a Misiurewicz parameter, the spacetime domain
for elementary cellular automaton rule 22, and finally the chaotic
crystallographic structure of a close-packed polytypic material---Zinc
Sulfide---as determined from experimental X-ray diffractograms.

%

\allowdisplaybreaks

\section{Even Process, a prototype sofic system}

\begin{figure}[h!]
  \centering
  \includegraphics[width=0.35\textwidth]{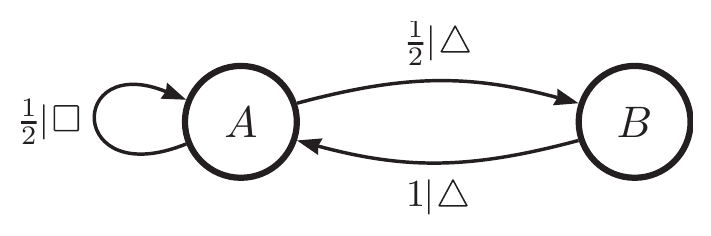}
\caption{An HMM $\mathcal{M}$ (in this case, the \eM) that generates the
  Even Process.
  }
\label{fig:EPeM}
\end{figure}

Consider the Even Process, the stochastic process generated over the
two-symbol alphabet $\mathcal{A} = \{ \square, \triangle \}$ by the hidden
Markov model $\mathcal{M}$ of Fig. \ref{fig:EPeM}. Though finitely
specified, being a strictly sofic system it produces sequences that have
arbitrarily long correlations; e.g., sequences containing blocks
$\square(\triangle\triangle)^k\square$, for $k = 0, 1, 2, \ldots$.

Ordering the causal states alphabetically, $\mathcal{M}$ has
symbol-labeled transition matrices:
\begin{align*}
T^{(\square)} =
	\begin{bmatrix}
	\sfrac{1}{2} 	& 0 \\
	0			& 0
	\end{bmatrix}
\text{and }
T^{(\triangle)} =
	\begin{bmatrix}
	0 		& \sfrac{1}{2} \\
	1		& 0
	\end{bmatrix}
  ~.
\end{align*}
From the state-transition matrix:
\begin{align}
T =
	\begin{bmatrix}
	\sfrac{1}{2} 	& \sfrac{1}{2} \\
	1	& 0
	\end{bmatrix}
	~,
\end{align}
we find the stationary distribution:
\begin{align}
\pi =
	\begin{bmatrix}
	\frac{2}{3}	& \frac{1}{3}
	\end{bmatrix} ~.
\label{eq: pi for ex HMM}
\end{align}

\begin{figure}[h!]
  \centering
  \includegraphics[width=0.45\textwidth]{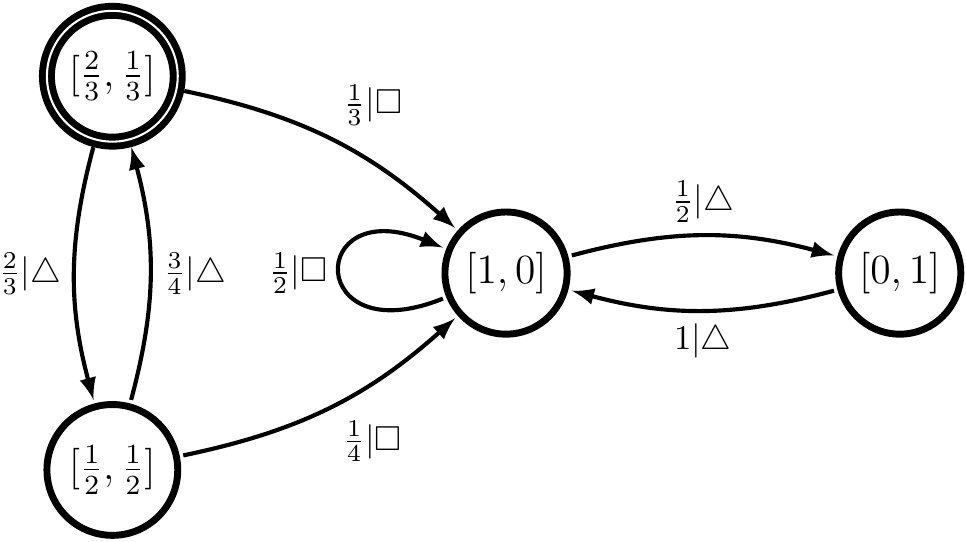}
\caption{$\mathcal{M}$'s mixed-state presentation $\mathcal{M}_\text{msp}$.
  The doubly-circled state denotes the start state $\mxstt{0} = \pi$. Inside
  each mixed state is a vector $[\Pr(A), \Pr(B)]$ which describes the induced
  distribution over the states of $\mathcal{M}$.
  }
\label{fig:Mmsp}
\end{figure}

The mixed-state presentation $\mathcal{M}_\text{msp}$ gives the dynamics over
$\mathcal{M}$'s state distributions, starting from the stationary
distribution $\langle \pi |$, induced by observed words $w \in \Abet^*$.
$\mathcal{M}_\text{msp}$ can be algorithmically generated from $\mathcal{M}$.
A state transition diagram for $\mathcal{M}_\text{msp}$ is given in Fig.
\ref{fig:Mmsp}. $\mathcal{M}_\text{msp}$ has state-transition matrix:
\begin{align}
W = \sum_{x \in \mathcal{A}} W^{(x)}  =
\begin{bmatrix}
0 			& \sfrac{2}{3} 	& \sfrac{1}{3}  		& 0 			\\
\sfrac{3}{4} 	& 0 			& \sfrac{1}{4} 		& 0 			\\
0			& 0 			& \sfrac{1}{2} 		& \sfrac{1}{2} 	\\
0			& 0 			& 1  				& 0
\end{bmatrix} .
\end{align}

Solving $\det (\lambda I - W) = 0$ gives $W$'s eigenvalues:
\begin{align*}
\Lambda_W &=
	\left\{
	1, \,
	\tfrac{\sqrt{2}}{2}, \,
	- \tfrac{\sqrt{2}}{2}, \,
	- \tfrac{1}{2}
  	\right\}
    ~.
\end{align*}
For each, we find the corresponding projection operator $W_\lambda$ via:
\begin{align*}
\label{eq: W_lambda algorithm}
W_{\lambda} & = \prod_{\substack{\zeta \in \Lambda_W \\ \zeta \neq \lambda }}
  \frac{W - \zeta I }{\lambda  - \zeta}
  ~,
\end{align*}
obtaining:
\begin{align*}
W_1
& =
\begin{bmatrix}
0 		& 0 		& \sfrac{2}{3}  		& \sfrac{1}{3} 			\\
0 		& 0 		& \sfrac{2}{3}  		& \sfrac{1}{3} 			\\
0 		& 0 		& \sfrac{2}{3}  		& \sfrac{1}{3} 			\\
0 		& 0 		& \sfrac{2}{3}  		& \sfrac{1}{3}
\end{bmatrix}
~, \\
W_{\sfrac{\sqrt{2}}{2}}
& =
\begin{bmatrix}
\sfrac{1}{2} 		& \phantom{-} \sfrac{\sqrt{2}}{3} 		& - \sfrac{(2 + \sqrt{2})}{6}  		& - \sfrac{(1 + \sqrt{2})}{6} 	\\
\phantom{-} \sfrac{3 \sqrt{2}}{8} 		& \sfrac{1}{2} 		& - \sfrac{(1 + \sqrt{2})}{4} 		& - \sfrac{(2 + \sqrt{2})}{8}	\\
0 		& 0 		& 0 		& 0 			\\
0 		& 0 		& 0 		& 0
\end{bmatrix}
~, \\
W_{- \sfrac{\sqrt{2}}{2}}
& =
\begin{bmatrix}
\sfrac{1}{2} 		& - \sfrac{\sqrt{2}}{3} 		& - \sfrac{(2 - \sqrt{2})}{6}  		& - \sfrac{(1 - \sqrt{2})}{6} 	\\
- \sfrac{3 \sqrt{2}}{8} 		& \sfrac{1}{2} 		& - \sfrac{(1 - \sqrt{2})}{4} 		& - \sfrac{(2 - \sqrt{2})}{8}	\\
0 		& 0 		& 0 		& 0 			\\
0 		& 0 		& 0 		& 0
\end{bmatrix}
~, \\
\intertext{and }
W_{-\sfrac{1}{2}}
& =
\begin{bmatrix}
0 		& 0 		& \phantom{-}0  		& \phantom{-}0 			\\
0 		& 0 		& - \sfrac{1}{6}  		& \phantom{-}\sfrac{1}{6} 			\\
0 		& 0 		& \phantom{-}\sfrac{1}{3}  		& - \sfrac{1}{3} 			\\
0 		& 0 		& - \sfrac{2}{3}  		& \phantom{-}\sfrac{2}{3}
\end{bmatrix}
~.
\end{align*}
Note that $W_1 = \ket{\one} \bra{\pi_W}$, which is always the case for an ergodic process.

We construct $\StartMS$ by placing all of the initial mass at
$\mathcal{M}_\text{msp}$'s start state, representing the stationary
distribution $\pi$ over the original presentation $\mathcal{M}$:
\begin{align*}
\StartMS  &=
  \begin{bmatrix}
   1 &  0 & 0 & 0
  \end{bmatrix}
  ~ .
\end{align*}

Table \ref{tab:MuWEtc} collects together the spectral
quantities of $W$ necessary for the exact calculation of the various complexity measures.

Different measures of complexity track the evolution of different types of information in (or about) the system.
The entropy of transitioning from the various states of uncertainty is given by the ket $\ket{H(W^\Abet)}$, whereas the internal entropy of the states of uncertainty themselves is given by the ket $\ket{H \! \left[ \eta \right]}$.
From the labeled transition matrices of the mixed-state presentation, we find:
\begin{align*}
\ket{H(W^\Abet)} & =
	\begin{bmatrix}
	\log_2(3) - \sfrac{2}{3}  \\
	2 - \frac{3}{4} \log_2(3) \\
	1 \\
	0
	\end{bmatrix} .
\end{align*}
And from the mixed states themselves,
\begin{align*}
\sum_{\eta \in \MxSSet} \eta \ket{\delta_\eta} & =
	\begin{bmatrix}
	( \sfrac{2}{3} \, , \;  \sfrac{1}{3} ) \\
	( \sfrac{1}{2} \, , \; \sfrac{1}{2} ) \\
	( 1\, , \; 0 ) \\
	( 0 \, , \; 1 )
\end{bmatrix} ,
\end{align*}
we have
\begin{align*}
\ket{H \! \left[ \eta \right]} & =
	\begin{bmatrix}
	\log_2(3) - \sfrac{2}{3}  \\
	1 \\
	0 \\
	0
\end{bmatrix} .
\end{align*}

\begin{table}
  \centering
\begin{tabular}{rrl}
  \hline
$\lambda$  & 	$\;\;\;\; \nu_\lambda$  & 	$\;\;\;\;\;\;\;\;\;\;\;\;\;\;\;\;\;\;\;\;
\StartMS W_{\lambda} $	\\
  \hline \\
  $1$ 	& 1			& \;\;\;\;\;\;$[ 0 \;\;\;\;\;\;\;\; 0 \;\;\;\;\;\;\;\;\;\;\; \tfrac{2}{3} \;\;\;\;\;\;\;\;\;\;\; \tfrac{1}{3} ]$ 	\\ \\ 
  $\sfrac{\sqrt{2}}{2}$    & 1		& \;\;\;\;\;\;$[ \tfrac{1}{2} \;\;\;\;\; \tfrac{\sqrt{2}}{3} \;\;\;\;\;\;  \tfrac{-2- \sqrt{2}}{6} \;\;\;\;\;  \tfrac{- 1 - \sqrt{2}}{6} ]$ 	\\ \\  
  $-\sfrac{\sqrt{2}}{2}$    & 1		& \;\;\;\;\;\;$[ \tfrac{1}{2} \;\;\;\;  \tfrac{- \sqrt{2}}{3} \;\;\;\;\;  \tfrac{-2+ \sqrt{2}}{6} \;\;\;\;\; \tfrac{- 1 + \sqrt{2}}{6} ]$ 	\\ \\ 
  $-\sfrac{1}{2}$    	& 1		& \;\;\;\;\;\;$[ 0 \;\;\;\;\;\;\;\; 0 \;\;\;\;\;\;\;\;\;\;\; 0 \;\;\;\;\;\;\;\;\;\;\;\, 0 ]$ 	\\ 
  \hline
\end{tabular}
\caption{Useful spectral quantities for the Even Process.}
\label{tab:MuWEtc}
\end{table}

\begin{figure}[h!]
  \centering
  \includegraphics[width=\columnwidth]{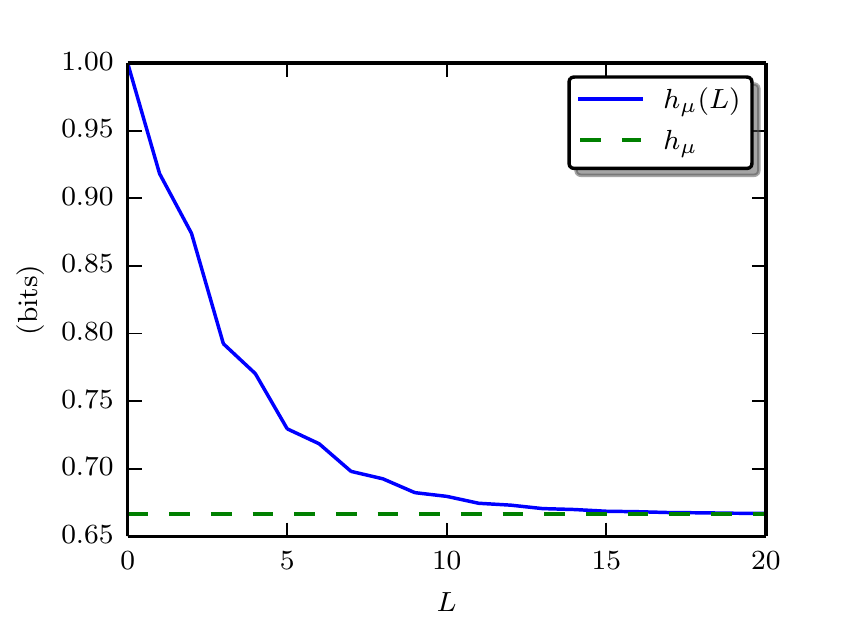}
  \caption{The entropy-rate convergence $\hmu(L)$ for the Even Process as a
  function of $L$, imposing the boundary condition that
  $\hmu(0) = \log_2 |\MeasAlphabet| = 1$.
  }
\label{fig:hL4EvenProcess}
\end{figure}

Hence, the finite-$L$ entropy-rate convergence for $L \geq 1$ is given by:
\begin{align*}
h_\mu(L) & =
  \sum_{\lambda \in \Lambda_W} \lambda^{L-1}
  \StartMS  W_{\lambda}  | H(W^\Abet) \rangle \\
& =
\begin{cases}
\tfrac{2}{3}
+ \left(\tfrac{\sqrt{2}}{2}\right)^{L-1} \left(  - \tfrac{\sqrt{2}}{2} \log_2(3) + \sqrt{2}  \right)
& \text{for even } L \\
\tfrac{2}{3}
+ \left(\tfrac{\sqrt{2}}{2}\right)^{L-1} \left(  \log_2(3)  - \tfrac{4}{3}   \right)
& \text{for odd } L
\end{cases}
  .
\end{align*}
This function is shown in Fig. \ref{fig:hL4EvenProcess}.
Similarly, the average state uncertainty after $L$ observations is given by:
\begin{align*}
\mathcal{H}(L) & =
  \sum_{\lambda \in \Lambda_W} \lambda^L
  \StartMS  W_{\lambda}  | H[\mxst] \rangle \\
& =
\left( \tfrac{\sqrt{2}}{2} \right)^L
	\biggl[ \tfrac{1}{2} \log_2(3) - \tfrac{1}{3} + \tfrac{\sqrt{2}}{3} \biggr. \\
& \qquad
	    \biggl. + (-1)^L \left( \tfrac{1}{2} \log_2(3) - \tfrac{1}{3} - \tfrac{\sqrt{2}}{3} \right) \biggr] \\
& =
\begin{cases}
\left( \log_2(3) - \sfrac{2}{3} \right)  \left(\tfrac{\sqrt{2}}{2}\right)^{L}  
& \text{for even } L \\
\tfrac{2 \sqrt{2} }{3} \left(\tfrac{\sqrt{2}}{2}\right)^{L}  
& \text{for odd } L
\end{cases}
  ~.
\end{align*}

For the scalar complexity measures, we find:
\begin{align*}
\hmu &= 2/3 ~\text{bits per step} , \\
\Cmu &= \log_2(3)  - \tfrac{2}{3}  ~\text{bits} ,  \\
\EE &= \log_2(3)  - \tfrac{2}{3}  ~\text{bits} ,  \\
\TI & = 2 \log_2(3) ~\text{bits-symbols} , \text{ and } \\
\SI & = 2 \log_2(3) ~\text{bits} .
\end{align*}

%

\section{Misiurewicz Tent Map: Intrinsic Computation in a Continuous-state Chaotic Dynamical System}

\begin{figure}[h!]
  \centering
  \includegraphics[width=0.47\textwidth]{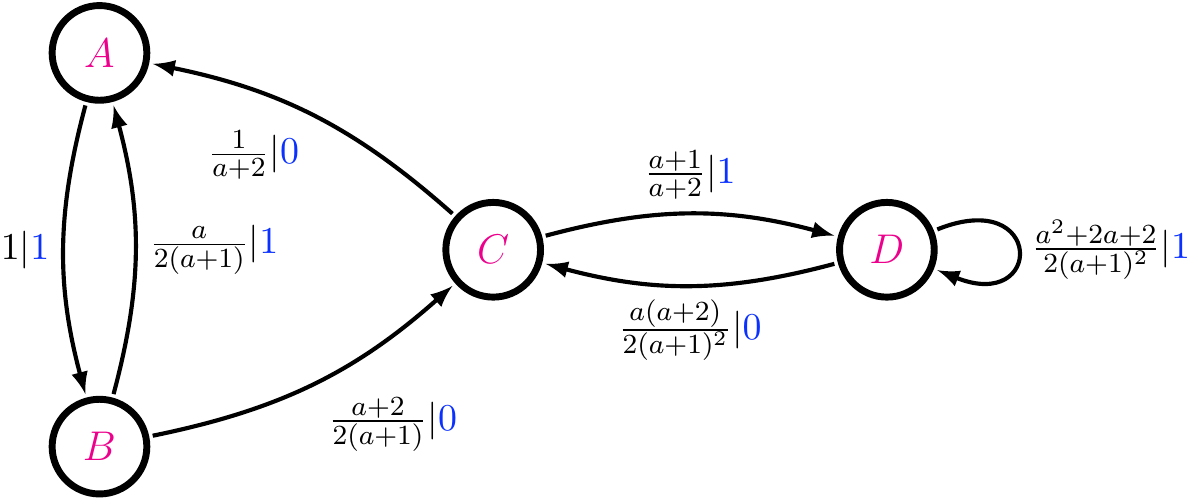} 	
\caption{The \eM\ $\mathcal{M}$ that captures the binary symbolic dynamics at
  the chosen Misiurewicz parameter. The transitions are labeled $p|x$ where
  $p$ is the transition probability and
  $x \in \{{\color{blue} 0},{\color{blue} 1}\}$ is the emitted symbol.
  Adapted from Ref. [A1] with permission.
  }
\label{fig:TentMapeM}
\end{figure}

The Tent Map of the unit interval:
\begin{align*}
x_{n+1} = a \cdot \min \{x_n,1-x_n\} ~,
\end{align*}
for $x_0 \in [0,1]$, is a well studied chaotic dynamical system. We set the
parameter $a$ to one of the so-called Misiurewicz values:
\begin{align*}
a & = \alpha + \tfrac{2}{3 \alpha} \\
  & \approx 1.76929235 ~,
\end{align*}
where $\alpha = \sqrt[3]{\sqrt{19/27} + 1}$. At this setting, the iterate of
the map's maximum ($x_c = 1/2$) becomes period-$1$ after the third iterate.
For a detailed analysis of this map see Ref. [A1]. Here, we use the
\eM\ found there for the binary symbolic dynamics observed with a generating
partition that divides the interval at $x_c$; see Fig. \ref{fig:TentMapeM}.

$\mathcal{M}$ has the two-symbol alphabet $\mathcal{A} = \{ 0, 1 \}$ with the
corresponding symbol-labeled transition matrices:
\begin{align*}
T^{(0)} &=
	\begin{bmatrix}
	0 & 0 & 0 & 0 \\
	0 & 0 & \frac{a+2}{2(a+1)} & 0 \\
	\frac{1}{\phantom{2}a+2 \phantom{()}} & 0 & 0 & 0 \\
	0 & 0 & \frac{a(a+2)}{2(a+1)^2} & 0 \\
	\end{bmatrix}
\text{and } \\
T^{(1)} &=
	\begin{bmatrix}
	0 & 1 & 0 & 0 \\
	\frac{a}{2(a+1)} & 0 & 0 & 0 \\
	0 & 0 & 0 & \frac{a+1}{a+2} \\
	0 & 0 & 0 & \frac{a^2+2a+2}{2(a+1)^2} \\
	\end{bmatrix}
  ~.
\end{align*}
We find the stationary distribution from the state-transition matrix $T$:
\begin{align}
\bra{\pi} = \begin{bmatrix} \pi_A & \pi_B & \pi_C & \pi_D \end{bmatrix}~,
\end{align}
with
\begin{align}
    \pi_A &= \pi_B = \frac{2(a+a^2)}{2+14a+14a^2+3a^3}\\
    \pi_C &= \frac{4a+4a^2+a^3}{2+14a+14a^2+3a^3}\\
    \pi_D &= \frac{2(1 + 3a + 3a^2 + a^3)}{2+14a+14a^2+3a^3}~.
\end{align}
So,
\begin{align*}
\bra{\eta_\varnothing} & =
\bra{\pi} \\
	& \propto \begin{bmatrix} 2a(a+1) & 2a(a+1) & a(a+2)^2 & 2(a+1)^3 \end{bmatrix}~, \\
\bra{\eta_0} &
	\propto \bra{\eta_\varnothing} T^{(0)} \\
	& \propto \begin{bmatrix} 1 & 0 & a+2 & 0 \end{bmatrix}~, \\
\bra{\eta_1} &
	\propto \bra{\eta_\varnothing} T^{(1)} \\
	& \propto \begin{bmatrix} a^2 & 2a(a+1) & 0 & 2(a+1)^3 \end{bmatrix}~, \\
\bra{\eta_{00}} &
	\propto \bra{\eta_0} T^{(0)} \\
	& \propto \begin{bmatrix} 1 & 0 & 0 & 0 \end{bmatrix}~, \\
\bra{\eta_{01}} &
	\propto \bra{\eta_0} T^{(1)} \\
	& \propto \begin{bmatrix} 0 & 1 & 0 & a+1 \end{bmatrix}~, \\
\bra{\eta_{10}} &
	\propto \bra{\eta_1} T^{(0)} \\
	& \propto \begin{bmatrix} 0 & 0 & 1 & 0 \end{bmatrix}~, \\
%
\bra{\eta_{11}} &
	\propto \bra{\eta_1} T^{(1)} \\
	& \propto \begin{bmatrix} a^2 & a^2 & 0 & (a+1)(a^2 + 2a + 2) \end{bmatrix}~,   \\
\bra{\eta_{001}} &
	\propto \bra{\eta_{00}} T^{(1)} \\
	& \propto \begin{bmatrix} 0 & 1 & 0 & 0 \end{bmatrix}~, \\
\bra{\eta_{011}} &
	\propto \bra{\eta_{01}} T^{(1)} \\
	& \propto \begin{bmatrix} a & 0 & 0 & a^2+2a+2 \end{bmatrix}~, \\
\intertext{and }
\bra{\eta_{101}} &
	\propto \bra{\eta_{10}} T^{(1)} \\
	& \propto \begin{bmatrix} 0 & 0 & 0 & 1 \end{bmatrix}~,
\end{align*}
give all of the distinct mixed states, labeled in subscripts by the first
shortest word that induces the mixed state from the initial mixed state
$\eta_\varnothing = \pi$.

\begin{figure}[h!]
  \centering
  \includegraphics[width=0.49\textwidth]{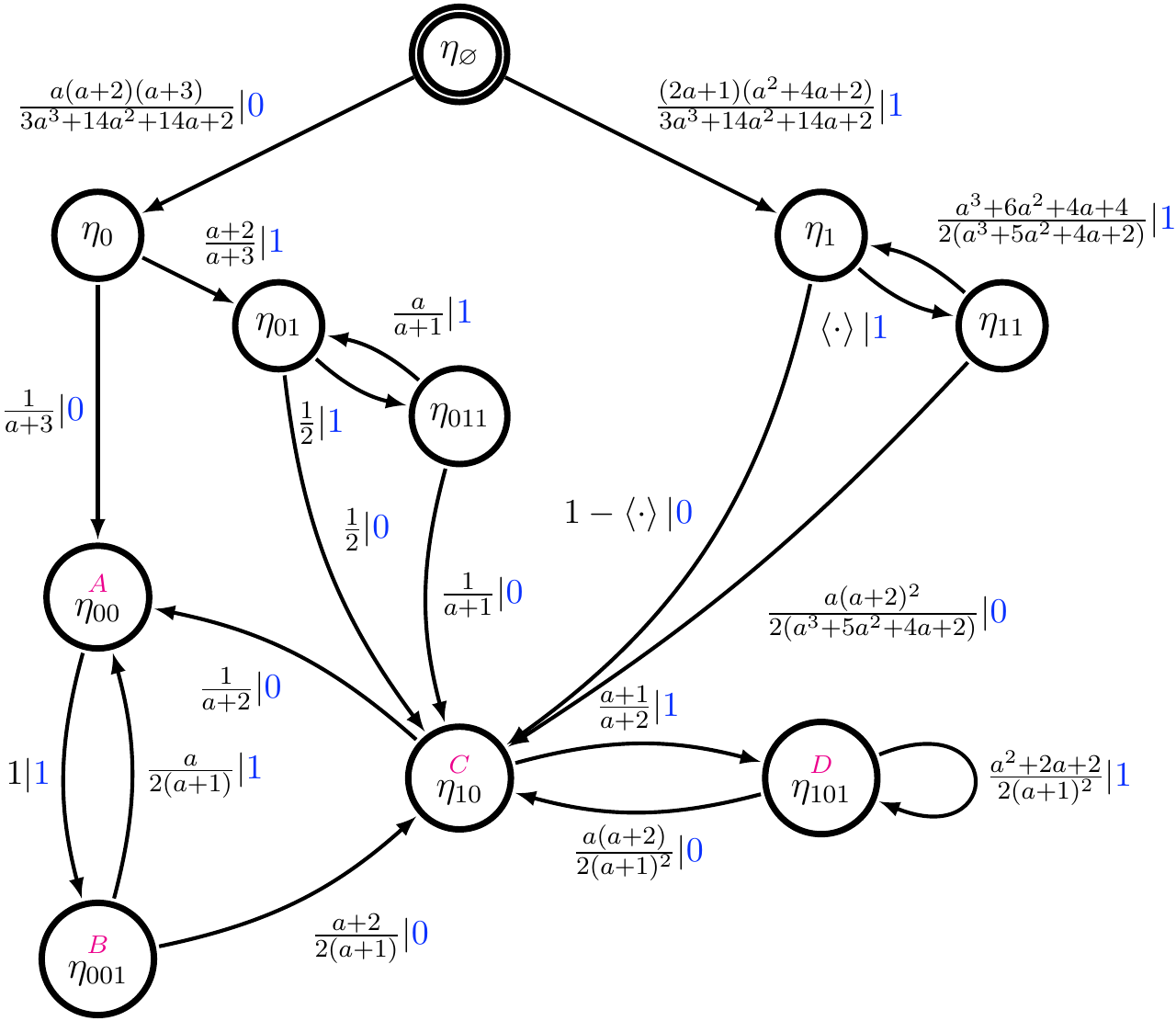}
\caption{The mixed-state presentation $\mathcal{M}_\text{msp}$ of the symbolic
  dynamics of the Tent Map at a Misiurewicz parameter. Due to space constraints
  we represent $W_{\eta_{1} \to \eta_{11}} = \frac{a^3 + 5a^2 + 4a + 2}{(2a+1)(a^2+4a+2)}$ as $\braket{\cdot}$ and $W_{\eta_{1} \to \eta_{10}} = \frac{a(a+2)^2}{(2a+1)(a^2+4a+2)}$ as $1 - \braket{\cdot}$.
  }
\label{fig:TentMSP}
\end{figure}

The mixed-state presentation is shown in Fig. \ref{fig:TentMSP}.
Keying off the graph's topology, we order the mixed states as:
\begin{align*}
\MxSSet = \{ \pi, \; \eta_0, \; \eta_{01}, \eta_{011}, \;
			 \eta_{1}, \eta_{11}, \;
  \overset{{\color{magenta} A}}{\eta_{00}},
  \overset{{\color{magenta} B}}{\eta_{001}},
  \overset{{\color{magenta} C}}{\eta_{10}},
  \overset{{\color{magenta} D}}{\eta_{101}}
  \},
\end{align*}
which is the ordering we use for the transition matrix and the bras and kets.
We put the \eM's recurrent state names---$A$, $B$, $C$, and $D$---above the
last four mixed states since they are isomorphic. That is, the recurrent states
of the mixed-state presentation (of the \eM) are effectively the recurrent
states of the \eM\ endowed with peaked distributions that uniquely identify
themselves among the causal states.

The transition probabilities between mixed states can be calculated via:
\begin{align*}
W_{\eta_\varnothing \to \eta_0} =
	\frac{\bra{\eta_\varnothing} T^{(0)} \ket{\one}}{\bra{\eta_\varnothing} \one \rangle }
	& = \frac{a(a+2)(a+3)}{3a^3+14a^2+14a+2} \\
	& \approx 0.364704 \\
W_{\eta_\varnothing \to \eta_1} =
	\frac{\bra{\eta_\varnothing} T^{(1)} \ket{\one}}{\bra{\eta_\varnothing} \one \rangle }
	& = \frac{2a^3 + 9a^2 + 8a + 2}{3a^3+14a^2+14a+2} \\
	& \approx 0.635296 \\
W_{\eta_0 \to \eta_{00}} =
	\frac{\bra{\eta_{0}} T^{(0)} \ket{\one}}{\bra{\eta_0} \one \rangle }
	& = \frac{1}{a+3} \\
	& \approx 0.209675 \\
\end{align*}
and so on, to determine the transition matrix for the mixed-state presentation:
\begin{widetext}
\begin{align*}
&W = \sum_{x \in \mathcal{A}} W^{(x)} =
\begin{bmatrix}
0 & 		\frac{a(a+2)(a+3)}{3a^3+14a^2+14a+2} & 	0 & 		0& 		\frac{2a^3 + 9a^2 + 8a + 2}{3a^3+14a^2+14a+2} & 		0 &
		0 & 			0 & 		0 & 		0 \\
0 & 		0 & 			\frac{a+2}{a+3} & 	0& 		0 & 		0 &
		\frac{1}{a+3} & 			0 & 		0 & 		0 \\
0 & 		0 & 			0 & 		\frac{1}{2}& 		0 & 		0 &
		0 & 			0 & 		\frac{1}{2} & 		0 \\
0 & 		0 & 			\frac{a}{a+1} & 		0& 		0 & 		0 &
		0 & 			0 & 		\frac{1}{a+1} & 		0 \\
0 & 		0 & 			0 & 		0& 		0 & 		\frac{a^3 + 5a^2 + 4a + 2}{2a^3+9a^2+8a+2} &
		0 & 			0 & 		\frac{a(a+2)^2}{2a^3+9a^2+8a+2} & 		0 \\
0 & 		0 & 			0 & 		0& 		\frac{a^3 + 6a^2 + 4a + 4}{2(a^3+5a^2+4a+2)} & 		0 &
		0 & 			0 & 		\frac{a(a+2)^2}{2(a^3+5a^2+4a+2)} & 		0 \\
0 & 		0 & 			0 & 		0& 		0 & 		0 &
		0 & 			1 & 		0 & 		0 \\
0 & 		0 & 			0 & 		0& 		0 & 		0 &
		\frac{a}{2(a+1)} & 0 & 	\frac{a+2}{2(a+1)} & 	0 \\
0 & 		0 & 			0 & 		0& 		0 & 		0 &
		\frac{1}{\phantom{2}a+2} & 0 & 0 & 	\frac{a+1}{a+2} \\
0 & 		0 & 			0 & 		0& 		0 & 		0 &
		0 & 			0 & 		\frac{a(a+2)}{2(a+1)^2} & 	\frac{a^2+2a+2}{2(a+1)^2} \\
\end{bmatrix} ~.
\end{align*}
\end{widetext}

The mixed states themselves can be cast as the column vector:
\begin{align*}
\sum_{\eta \in \MxSSet} \eta \ket{\delta_\eta} & =
	\begin{bmatrix}
	\tfrac{2(a+1)}{3a^3+14a^2+14a+2} \bigl( a,  a,  \tfrac{a(a+2)^2}{2(a+1)},  (a+1)^2 \bigr) \\
	\tfrac{1}{a+3} \bigl( 1, 0, a+2, 0 \bigr) \\
	\tfrac{1}{a+2} \bigl( 0, 1, 0, a+1 \bigr) \\
	\tfrac{1}{(a+1)(a+2)} \bigl( a, 0, 0, \frac{2(a+1)^2}{a} \bigr) \\   
	\tfrac{1}{2a^3+9a^2+8a+2} \bigl( a^2,  2a(a+1),  0,  2(a+1)^3 \bigr) \\
	\tfrac{1}{a^3+5a^2+4a+2} \bigl( a^2,  a^2,  0,  \frac{2(a+1)^3}{a} \bigr) \\  
	\bigl( 1, 0, 0, 0 \bigr) \\
	\bigl( 0, 1, 0, 0 \bigr) \\
	\bigl( 0, 0, 1, 0 \bigr) \\
	\bigl( 0, 0, 0, 1 \bigr)
\end{bmatrix} ,
\end{align*}
from which we obtain the internal entropy of the mixed states. This is perhaps
best left represented as:
\begin{align*}
\ket{H[\eta]}
&=
\sum_{\eta \in \MxSSet}  \ket{\delta_\eta} \times - \sum_{\st \in \SSet} \bra{\delta_\st} \eta \rangle \log_2 \bra{\delta_\st} \eta \rangle \\
& =
\begin{bmatrix}
	H[\pi] \\
	H[\eta_0] \\
	H[\eta_{01}] \\
	H[\eta_{011}] \\
	H[\eta_{1}] \\
	H[\eta_{11}] \\
	0 \\ 		
	0 \\		
	0 \\		
	0		
\end{bmatrix}
\approx
\begin{bmatrix}
	1.731517488310359 \\
	0.740860503264943 \\
	0.834641915284059 \\
	0.656560846029285 \\
	0.970187276943296 \\
	0.942259650948105 \\
	0 \\ 		
	0 \\		
	0 \\		
	0		
\end{bmatrix} \text{ bits}.
\end{align*}
Evidently, the internal entropy of the recurrent mixed states (the last four rows) is zero---as expected of the recurrent mixed states of the \eM.
From $\Cmu = \StartMS H[\eta] \rangle$, the statistical complexity
of the process is $\Cmu = H[\pi]$.

Let $Q$ be the $6 \times 6$ substochastic matrix of transitions among the ordered set of transient mixed states $\MxSSet_Q = \{ \pi, \; \eta_0, \; \eta_{01}, \eta_{011}, \; \eta_{1}, \eta_{11} \}$.
Then, for some $6 \times 4$ matrix $B$, we write:
\begin{align*}
W & =
  \begin{bmatrix}
    Q & B \\
    0 & T
  \end{bmatrix} ,
\end{align*}
where the `0' just above is the $4 \times 6$ matrix of all zeros. We see that
to obtain the synchronization information, we need only powers of $Q$.
$B$ and $T$ are irrelevant to calculating $\SI$, once $Q$ has been obtained:
\begin{align*}
\SI & = \sum_{L = 0}^{\infty} \StartMS W^{L} \ket{H[\mxst]} \\
  & = \sum_{L = 0}^{\infty}
    \begin{bmatrix}
        \StartMS & 0
    \end{bmatrix}
    \begin{bmatrix}
        Q^L & (\cdot) \\
        0 & T^L
    \end{bmatrix}
    \begin{bmatrix}
        \ket{H[\mxst \in \MxSSet_Q]} \\
        0
    \end{bmatrix}     \\
& =
\sum_{L = 0}^{\infty} \StartMS Q^{L} \ket{H[\mxst \in \MxSSet_Q]} ,
\end{align*}
where the zeros that appear above inherit the appropriate dimensions for matrix
multiplication. And, we reuse the notation $\StartMS$ to refer to the shortened
$1 \times 6$ row-vector with the last four zeros removed. We are now ready to
obtain the synchronization information via the spectral decomposition of $Q^L$.

From:
\begin{align*}
\text{det}(\lambda I - Q)
&= \lambda^2 \left(\lambda^2 - \tfrac{a}{2(a+1)} \right) \left( \lambda^2 - \tfrac{a^3 + 6a^2 + 4a + 4}{2(2a^3 + 9a^2 + 8a + 2)}  \right) \\
& = \lambda^2 \left(\lambda^2 - \tfrac{a}{2(a+1)} \right)^2 ,
\end{align*}
we obtain $Q$'s eigenvalues:
\begin{align*}
\Lambda_Q
&=
\left\{ 0, \, \pm \sqrt{\tfrac{a}{2(a+1)}} \right\} ~.
\end{align*}

Since the index of the zero eigenvalue is greater than one ($\nu_0 = 2$,
since $a_0 = 2$ and $g_0 = 1$), $Q$ is not diagonalizable. Nevertheless,
since all eigenvalues besides the zero eigenvalue have index equal to unity,
we can calculate powers of $Q$ by the slightly more general spectral
decomposition given in Eq. (\ref{eq:GeneralSpectralDecomp}):
\begin{align}
Q^L
& =
\Biggl\{
	\sum_{\lambda \in \Lambda_Q}
	\lambda^L  Q_\lambda
\Biggr\}
+
\sum_{N=1}^{\nu_0 - 1}  \delta_{L, N}  Q_0 Q^N \nonumber \\
& =
\label{eq: decomp of powers of Q}
\Biggl\{
	\sum_{\lambda \in \Lambda_Q}
	\lambda^L  Q_\lambda
\Biggr\}
+
\delta_{L, 1}  Q_0 Q ~.
\end{align}

Using Eq.\ \eqref{eq: decomp of powers of Q},
the synchronization information becomes:
\begin{align}
\label{eq:SIFormula4AlmostDiagQ}
\SI
& =
\StartMS Q_{0} Q \ket{H[\mxst \in \MxSSet_Q]} \nonumber \\
  & \qquad +
   \sum_{\lambda \in \Lambda_Q} \frac{1}{1-\lambda} \StartMS Q_\lambda \ket{H[\mxst \in \MxSSet_Q]}  ~.
\end{align}

Since the two eigenvalues besides the zero eigenvalue have index equal to unity, their projection operators can be obtained via:
\begin{align*}
Q_\lambda
& =
\left( \frac{Q}{\lambda} \right)^{\nu_0}  \negthickspace
  \prod_{\zeta \in \Lambda_Q \setminus \{ 0 \} \atop \zeta \neq \lambda}
    \frac{Q - \zeta I}{\lambda - \zeta} \\
& =
\left( \frac{Q}{\lambda} \right)^{2}
    \frac{Q +\lambda I}{2 \lambda} \\
& =
\tfrac{a+1}{a} Q^2 \left( \lambda^{-1} Q + I \right)
\end{align*}
for each $\lambda \in  \Bigl\{ \pm \sqrt{\tfrac{a}{2(a+1)}} \Bigr\}$. 
Then, since the projection operators must sum to the identity matrix,
$Q_0$ can be obtained via:
\begin{align*}
Q_0
&=
I - \sum_{\lambda \in \Lambda_Q \setminus \{ 0 \} } Q_\lambda \\
& = I - \tfrac{a+1}{a} Q^2 \left( \sqrt{\tfrac{2(a+1)}{a}} Q - \sqrt{\tfrac{2(a+1)}{a}} Q + 2 I \right) \\
& = I - \tfrac{2(a+1)}{a} Q^2 ~.
\end{align*}

Preparing to calculate $\SI$, we find:
\begin{widetext}
\begin{align*}
\sum_{\lambda \in \Lambda_Q \setminus \{ 0 \} } &
  \frac{1}{1-\lambda} \StartMS Q_\lambda \\
& =
\tfrac{a+1}{a}
		\StartMS \left\{
			\biggl( \tfrac{1}{\sqrt{\tfrac{a}{2(a+1)}} - \tfrac{a}{2(a+1)} } \biggr) Q^3 +
			\biggl( \tfrac{1}{1 - \sqrt{\tfrac{a}{2(a+1)}} } \biggr) Q^2
	 	+
			\biggl( \tfrac{-1}{\sqrt{\tfrac{a}{2(a+1)}} + \tfrac{a}{2(a+1)} } \biggr) Q^3 +
			\biggl( \tfrac{1}{1 + \sqrt{\tfrac{a}{2(a+1)}} } \biggr) Q^2
		\right\} \\
& =
\tfrac{a+1}{a}
		\StartMS \left\{
			\biggl( \tfrac{2}{1 - \tfrac{a}{2(a+1)} } \biggr) Q^3 +
			\biggl( \tfrac{2}{1 - \tfrac{a}{2(a+1)} } \biggr) Q^2
		\right\} \\
& =
\tfrac{4(a+1)^2}{a(a+2)} \StartMS \left( Q^3 + Q^2 \right) .
\end{align*}
\end{widetext}
Hence:
\begin{align*}
\sum_{\lambda \in \Lambda_Q } &
  \frac{1}{1-\lambda} \StartMS Q_\lambda \\
& =
\StartMS \left\{
			I +
			\tfrac{2(a+1)}{a}  \biggl( \tfrac{2(a+1)}{a+2} - 1 \biggr) Q^2 +
			\tfrac{4(a+1)^2}{a(a+2)} Q^3
		\right\} \\
& =
\StartMS \left(  
			I +
			\tfrac{2(a+1)}{a+2}  Q^2 +
			\tfrac{4(a+1)^2}{a(a+2)} Q^3
		\right)  
\end{align*}
and
\begin{align}
& \StartMS Q_0 Q +
\sum_{\lambda \in \Lambda_Q }
  \frac{1}{1-\lambda} \StartMS Q_\lambda \nonumber \\
& =
\StartMS \left(
			Q -
			\tfrac{2(a+1)}{a}  Q^3
		\right)  +
\StartMS \left(
			I +
			\tfrac{2(a+1)}{a+2}  Q^2 +
			\tfrac{4(a+1)^2}{a(a+2)} Q^3
		\right) \nonumber \\
& =
\label{eq:FirstRowOfQLDecomp4TentEx}
\StartMS \left\{ I + Q + \tfrac{2(a+1)}{a+2}  \left( Q^2 + Q^3 \right) \right\}
  ~.
\end{align}

From \eqref{eq:SIFormula4AlmostDiagQ} and \eqref{eq:FirstRowOfQLDecomp4TentEx}, we obtain the synchronization information,
\begin{align*}
\SI = & \StartMS
	\left( I + Q + \tfrac{2(a+1)}{a+2}  \left( Q^2 + Q^3 \right) \right)
	\ket{ H[\mxst \in \MxSSet_Q] } ~,
\end{align*}
which can be written more explicitly as:
\begin{align*}
\SI = H[\pi]
+&
\tfrac{1}{3a^3 + 14a^2 + 14a + 2}
    \biggl\{
        a(a+2)(a+3) H[\eta_0]
    \biggr. \\
    & 
    + a(a+1)(a+2) \Bigl( 2 H[\eta_{01}] + H[\eta_{011}] \Bigr)  \\
    & 
    +\tfrac{2(a+1)}{a+2}
       \Bigl( \tfrac{a+1}{a} (a^3 + 6a^2 + 4a + 4) H[\eta_{1}] \Bigr. \\
        &
        \biggl.
        \Bigl.
        \qquad \qquad \quad  + (a^3 + 5a^2 + 4a + 2) H[\eta_{11}]  \Bigr)
    \biggr\}  \\
& \approx 3.880442712215985 \text{ bits}.
\end{align*}

%

\section{Intrinsic Computation in Spacetime}

Here, we calculate the informational properties of one of elementary cellular
automaton (ECA) rule 22's domains---its dominant spacetime invariant sets. See
Ref. [A2] for a discussion of spacetime domain and particle analysis and
related structures for ECA 54. We mention only the minimal prerequisites
necessary for analyzing spacetime intrinsic computation.

\begin{figure}[h!]
  \centering
  \includegraphics{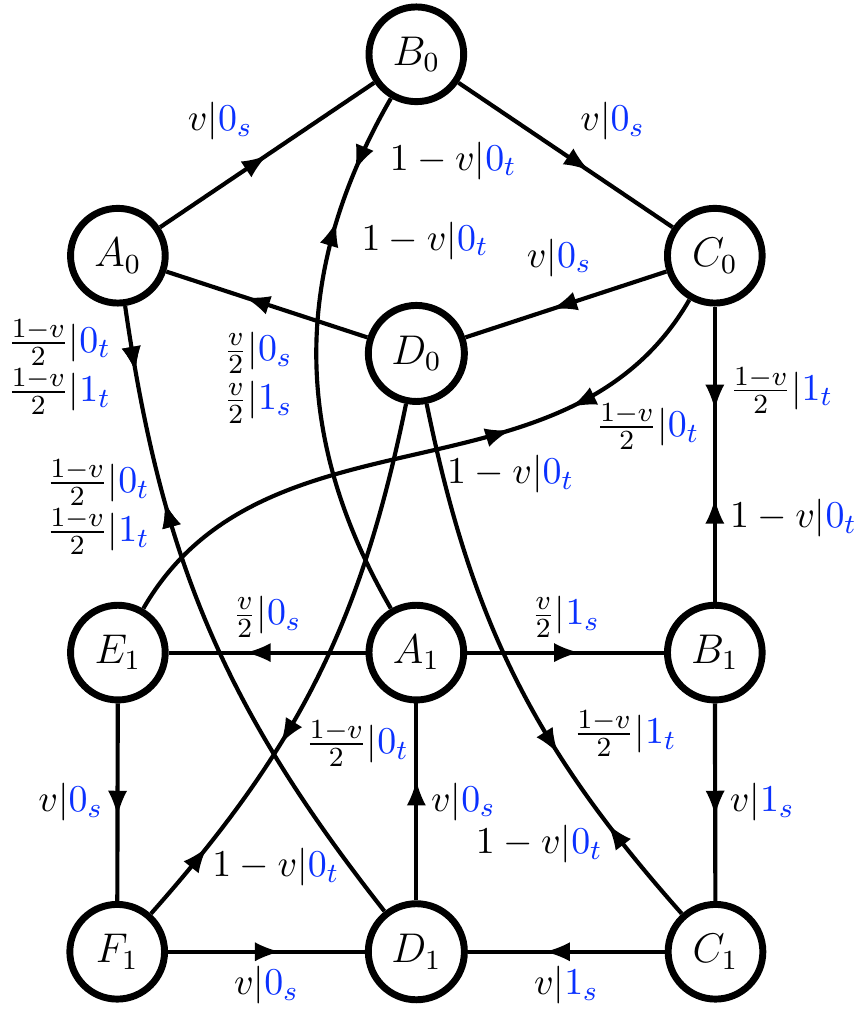}
\caption{The \eM\ $\mathcal{M}$ describing the spacetime configuration
  patches generated by ECA 22 that are spacetime shift invariant. When
  leaving a state, spacelike ($s$) and timelike ($t$) moves are made with
  probability dependent on the average ``velocity'' $v \in (0, 1)$. Here,
  $v = 0$ corresponds to a strictly timelike sequence of moves and $v = 1$
  corresponds to a strictly spacelike sequence of moves. Once a spacelike or
  timelike move has been chosen, if the topology of the \eM\ does not
  completely determine the next site value, then the site values ($0$ or $1$)
  occur with equal probability. The product of these form the transition
  probabilities listed on the edges. Note that the state-to-state transitions
  $D_0$ to $A_0$, $D_1$ to $A_0$, and $A_0$ to $D_1$ consist of two parallel
  transitions on $0$ and $1$, though they are depicted with a single edge in
  the diagram. Those edges do have a pair of labels as appropriate.
  }
\label{fig:ECA22DomaineM}
\end{figure}

ECA 22 generates a series of 1D spatial configurations. The domain is a
collection of spacetime patches consisting of binary values $\{0,1\}$ at each
site that is dynamically invariant under spatial and temporal shifts. It is
described by
the \eM\ shown in Fig.  \ref{fig:ECA22DomaineM} over the four-letter alphabet
$\ProcessAlphabet = \{0_s, 1_s, 0_t, 1_t \}$, where the site values are
subscripted by $s$ when the value is seen when making a \emph{spacelike}
move and by $t$ when the value is seen when making a \emph{timelike} move.
Thus, the \eM\ describes the set of binary strings observed when taking all
(right) spacelike and (positive) timelike moves across the spacetime lattice.

$\mathcal{M}$ has the symbol-labeled transition matrices:
\begin{align*}
T^{(0_s)} & =
	\left[
	\begin{array}{*{10}{>{\centering\arraybackslash$}p{0.4cm}<{$}}}
	  0  & v &  0  &  0  &  0  &  0  &  0  &  0  &  0  &  0  \\ 
	  0  &  0  & v &  0  &  0  &  0  &  0  &  0  &  0  &  0  \\ 
	  0  &  0  &  0  & v &  0  &  0  &  0  &  0  &  0  &  0  \\ 
	 v/2 &  0  &  0  &  0  &  0  &  0  &  0  &  0  &  0  &  0  \\ 
	  0  &  0  &  0  &  0  &  0  &  0  &  0  &  0  & v/2 &  0  \\ 
	  0  &  0  &  0  &  0  &  0  &  0  &  0  &  0  &  0  &  0  \\ 
	  0  &  0  &  0  &  0  &  0  &  0  &  0  &  0  &  0  &  0  \\ 
	  0  &  0  &  0  &  0  & v &  0  &  0  &  0  &  0  &  0  \\ 
	  0  &  0  &  0  &  0  &  0  &  0  &  0  &  0  &  0  & v \\ 
	  0  &  0  &  0  &  0  &  0  &  0  &  0  & v &  0  &  0  \\ 
	\end{array}
	\right]
\text{, } \\
T^{(1_s)} & =
	\left[
 	\begin{array}{*{10}{>{\centering\arraybackslash$}p{0.4cm}<{$}}}
	  0  &  0  &  0  &  0  &  0  &  0  &  0  &  0  &  0  &  0  \\ 
	  0  &  0  &  0  &  0  &  0  &  0  &  0  &  0  &  0  &  0  \\ 
	  0  &  0  &  0  &  0  &  0  &  0  &  0  &  0  &  0  &  0  \\ 
	 v/2 &  0  &  0  &  0  &  0  &  0  &  0  &  0  &  0  &  0  \\ 
	  0  &  0  &  0  &  0  &  0  & v/2 &  0  &  0  &  0  &  0  \\ 
	  0  &  0  &  0  &  0  &  0  &  0  & v &  0  &  0  &  0  \\ 
	  0  &  0  &  0  &  0  &  0  &  0  &  0  & v &  0  &  0  \\ 
	  0  &  0  &  0  &  0  &  0  &  0  &  0  &  0  &  0  &  0  \\ 
	  0  &  0  &  0  &  0  &  0  &  0  &  0  &  0  &  0  &  0  \\ 
	  0  &  0  &  0  &  0  &  0  &  0  &  0  &  0  &  0  &  0  \\ 
	\end{array}
	\right]
\text{, } \\
T^{(0_t)} & =
	\left[
	\begin{array}{*{10}{>{\centering\arraybackslash$}p{0.4cm}<{$}}}
	  0  &  0  &  0  &  0  &  0  &  0  &  0  & \tfrac{1-v}{2} &  0  &  0  \\ 
	  0  &  0  &  0  &  0  & {\scriptstyle 1-v} &  0  &  0  &  0  &  0  &  0  \\ 
	  0  &  0  &  0  &  0  &  0  &  0  &  0  &  0  & \tfrac{1-v}{2} &  0  \\ 
	  0  &  0  &  0  &  0  &  0  &  0  &  0  &  0  &  0  & \tfrac{1-v}{2} \\ 
	  0  & {\scriptstyle 1-v} &  0  &  0  &  0  &  0  &  0  &  0  &  0  &  0  \\ 
	  0  &  0  & {\scriptstyle 1-v} &  0  &  0  &  0  &  0  &  0  &  0  &  0  \\ 
	  0  &  0  &  0  & {\scriptstyle 1-v} &  0  &  0  &  0  &  0  &  0  &  0  \\ 
	 \tfrac{1-v}{2} &  0  &  0  &  0  &  0  &  0  &  0  &  0  &  0  &  0  \\ 
	  0  &  0  & {\scriptstyle 1-v} &  0  &  0  &  0  &  0  &  0  &  0  &  0  \\ 
	  0  &  0  &  0  & {\scriptstyle 1-v} &  0  &  0  &  0  &  0  &  0  &  0  \\ 
	\end{array}
	\right]
\text{and } \\
T^{(1_t)} & =
	\left[
	\begin{array}{*{10}{>{\centering\arraybackslash$}p{0.4cm}<{$}}}
	  0  &  0  &  0  &  0  &  0  &  0  &  0  & \tfrac{1-v}{2} &  0  &  0  \\ 
	  0  &  0  &  0  &  0  &  0  &  0  &  0  &  0  &  0  &  0  \\ 
	  0  &  0  &  0  &  0  &  0  & \tfrac{1-v}{2} &  0  &  0  &  0  &  0  \\ 
	  0  &  0  &  0  &  0  &  0  &  0  & \tfrac{1-v}{2} &  0  &  0  &  0  \\ 
	  0  &  0  &  0  &  0  &  0  &  0  &  0  &  0  &  0  &  0  \\ 
	  0  &  0  &  0  &  0  &  0  &  0  &  0  &  0  &  0  &  0  \\ 
	  0  &  0  &  0  &  0  &  0  &  0  &  0  &  0  &  0  &  0  \\ 
	 \tfrac{1-v}{2} &  0  &  0  &  0  &  0  &  0  &  0  &  0  &  0  &  0  \\ 
	  0  &  0  &  0  &  0  &  0  &  0  &  0  &  0  &  0  &  0  \\ 
	  0  &  0  &  0  &  0  &  0  &  0  &  0  &  0  &  0  &  0  \\ 
	\end{array}
	\right]
  ~.
\end{align*}
The state-transition matrix $T = \sum_{x \in \Abet} T^{(x)}$ is:
\begin{align*}
T & =
	\left[
	\begin{array}{*{10}{>{\centering\arraybackslash$}p{0.4cm}<{$}}}
	  0   & v   &  0  &  0  &  0   &  0      &  0      & {\scriptstyle 1-v} &  0  &  0  \\ 
	  0   &  0  & v   &  0  & {\scriptstyle 1-v} &  0      &  0      &  0  &  0   &  0  \\ 
	  0   &  0  &  0  & v   &  0   & \tfrac{1-v}{2} &  0   &  0  & \tfrac{1-v}{2} &  0  \\ 
	 v    &  0  &  0  &  0  &  0   &  0      & \tfrac{1-v}{2} &  0  &  0  & \tfrac{1-v}{2} \\ 
	  0   & {\scriptstyle 1-v} &  0 &  0  &  0  & v/2    &  0      &  0  & v/2 &  0  \\ 
	  0   &  0  & {\scriptstyle 1-v} &  0 &  0  &  0      &  v      &  0  &  0   &  0  \\ 
	  0   &  0  &  0  & {\scriptstyle 1-v} &  0 &  0      &  0      &  v  &  0   &  0  \\ 
	 {\scriptstyle 1-v} &  0  &  0  &  0  & v   &  0      &  0      &  0  &  0   &  0  \\ 
	  0   &  0  & {\scriptstyle 1-v} &  0  &  0 &  0      &  0      &  0  &  0   &  v  \\ 
	  0   &  0  &  0  & {\scriptstyle 1-v} &  0 &  0      &  0      &  v  &  0   &  0  \\ 
	\end{array}
	\right]
  ~.
\end{align*}
And, from it we find the stationary distribution:
\begin{align}
\bra{\pi} = \tfrac{1}{16}
  \begin{bmatrix} 2 & 2 & 2 & 2 & 2 & 1 & 1 & 2 & 1 & 1 \end{bmatrix} ~.
\end{align}

However, at the extremes of $v = 0$ and $v = 1$ the \eM\ breaks apart into an
ensemble of subprocesses. We analyze several subprocesses here, comparing the
complexity measures of path ensembles in space versus those in time.

\subsection{Timelike complexity}

\begin{figure}[h!]
  \centering
  \includegraphics[width=0.25\textwidth]{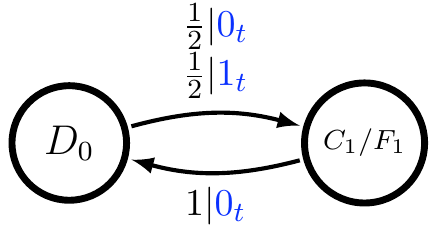}
\caption{The \eM\ of one of the strictly timelike subprocesses that appears at $v = 0$.
  }
\label{fig:ECA22StrictlyTimelikeSubprocess}
\end{figure}

For $v = 0$, one of the strictly timelike subprocesses collapses down to the noisy period-2 process shown in Fig.\ \ref{fig:ECA22StrictlyTimelikeSubprocess}.

\begin{figure}[h!]
  \centering
  \includegraphics[width=0.4\textwidth]{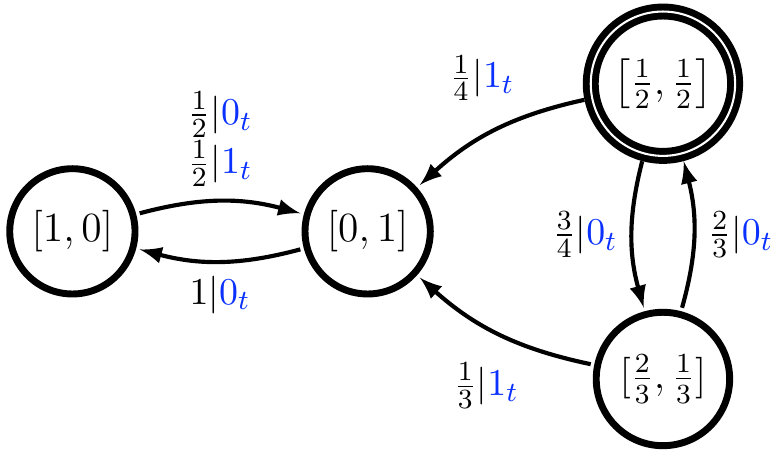}
\caption{The MSP of the strictly timelike subprocess shown in Fig.\ \ref{fig:ECA22StrictlyTimelikeSubprocess}.
  }
\label{fig:ECA22StrictlyTimelikeSubprocess_MSP}
\end{figure}

For this strictly timelike subprocess, we obtain the MSP shown in Fig.\ \ref{fig:ECA22StrictlyTimelikeSubprocess_MSP},
where the mixed states are labeled with their corresponding distribution
$\left[ \Pr(D_0), \Pr(C_1/F_1) \right] $.

If we order the set of mixed states as:
\begin{align*}
\MxSSet = \{ \pi = (\tfrac{1}{2}, \tfrac{1}{2}), \; \eta_0 = (\tfrac{2}{3}, \tfrac{1}{3}), \; \eta_{1} = (0, 1), \eta_{10} = (1, 0) \} ,
\end{align*}
then the mixed-state presentation has state-transition matrix:
\begin{align}
W = \sum_{x \in \mathcal{A}} W^{(x)}  =
\begin{bmatrix}
0 			& \sfrac{3}{4} 	& \sfrac{1}{4}  		& 0 		\\
\sfrac{2}{3} 	& 0 			& \sfrac{1}{3} 		& 0 		\\
0			& 0 			& 0 				& 1 		\\
0			& 0 			& 1  				& 0
\end{bmatrix} .
\end{align}

Solving $\det (\lambda I - W) = 0$ gives $W$'s eigenvalues:
\begin{align*}
\Lambda_W &=
	\left\{
	1, \,
	-1, \,
	\sqrt{\tfrac{1}{2}}, \,
	- \sqrt{\tfrac{1}{2}}
  	\right\}
    ~.
\end{align*}
For each, we find the corresponding projection operator $W_\lambda$ via:
\begin{align*}
W_{\lambda} & = \prod_{\substack{\zeta \in \Lambda_W \\ \zeta \neq \lambda }}
  \frac{W - \zeta I }{\lambda  - \zeta}
  ~,
\end{align*}
obtaining:
\begin{align*}
W_1
& =
- \tfrac{1}{2} I - \tfrac{1}{2} W + W^2 + W^3  \\
& =
\begin{bmatrix}
0 		& 0 		& \sfrac{1}{2}  		& \sfrac{1}{2} 			\\
0 		& 0 		& \sfrac{1}{2}  		& \sfrac{1}{2} 			\\
0 		& 0 		& \sfrac{1}{2}  		& \sfrac{1}{2} 			\\
0 		& 0 		& \sfrac{1}{2}  		& \sfrac{1}{2}
\end{bmatrix}
~, \\
W_{-1}
& =
- \tfrac{1}{2} I + \tfrac{1}{2} W + W^2 - W^3 \vphantom{a^{\bigl[ \bigr]}} \\
& =
\begin{bmatrix}
0 		& 0 		& \phantom{-}0  		& \phantom{-}0 			\\
0 		& 0 		& - \sfrac{1}{6}  		& \phantom{-}\sfrac{1}{6} 			\\
0 		& 0 		& \phantom{-}\sfrac{1}{2}  		& - \sfrac{1}{2} 			\\
0 		& 0 		& - \sfrac{1}{2}  		& \phantom{-}\sfrac{1}{2}
\end{bmatrix}
~, \\
W_{\sfrac{\sqrt{2}}{2}}
& =
I - W^2 + \sqrt{2} \left( W - W^3 \right) \vphantom{a^{\bigl[ \bigr]}}  \\
& =
\begin{bmatrix}
\sfrac{1}{2} 		& \phantom{-} \sfrac{3 \sqrt{2}}{8} 		& - \sfrac{(2 + \sqrt{2})}{8}  		& - \sfrac{(1 + \sqrt{2})}{4} 	\\
\phantom{-} \sfrac{\sqrt{2}}{3} 		& \sfrac{1}{2} 		& - \sfrac{(1 + \sqrt{2})}{6} 		& - \sfrac{(2 + \sqrt{2})}{6}	\\
0 		& 0 		& 0 		& 0 			\\
0 		& 0 		& 0 		& 0
\end{bmatrix}
~, \\
\intertext{and }
W_{- \sfrac{\sqrt{2}}{2}}
& =
I - W^2 - \sqrt{2} \left( W - W^3 \right)  \\
& =
\begin{bmatrix}
\sfrac{1}{2} 		& - \sfrac{3 \sqrt{2}}{8} 		& - \sfrac{(2 - \sqrt{2})}{8}  		& - \sfrac{(1 - \sqrt{2})}{4} 	\\
- \sfrac{\sqrt{2}}{3} 		& \sfrac{1}{2} 		& - \sfrac{(1 - \sqrt{2})}{6} 		& - \sfrac{(2 - \sqrt{2})}{6}	\\
0 		& 0 		& 0 		& 0 			\\
0 		& 0 		& 0 		& 0
\end{bmatrix}
~.
\end{align*}
Note that $W_1 = \ket{\one} \bra{\pi_W}$, again, since the timelike subprocess
is ergodic.

We construct $\StartMS$ by placing all of the initial mass at
$\mathcal{M}_\text{msp}$'s start state, representing the stationary
distribution $\pi$ over the original presentation $\mathcal{M}$:
\begin{align*}
\StartMS  &=
  \begin{bmatrix}
   1 &  0 & 0 & 0
  \end{bmatrix}
  ~ .
\end{align*}

Different measures of complexity track the evolution of different types of information in (or about) the system.
The entropy of transitioning from the various states of uncertainty is given by the ket $\ket{H(W^\Abet)}$, whereas the internal entropy of the states of uncertainty themselves is given by the ket $\ket{H \! \left[ \eta \right]}$.
From the labeled transition matrices of the mixed-state presentation, we find:
\begin{align*}
\ket{H(W^\Abet)} & =
	\begin{bmatrix}
	2 - \frac{3}{4} \log_2(3) \\
	\log_2(3) - \sfrac{2}{3}  \\
	0 \\
	1
	\end{bmatrix} .
\end{align*}
And from the mixed states themselves,
\begin{align*}
\sum_{\eta \in \MxSSet} \eta \ket{\delta_\eta} & =
	\begin{bmatrix}
	( \sfrac{1}{2} \, , \;  \sfrac{1}{2} ) \\
	( \sfrac{2}{3} \, , \; \sfrac{1}{3} ) \\
	( 0\, , \; 1 ) \\
	( 1 \, , \; 0 )
\end{bmatrix} ,
\end{align*}
we have
\begin{align*}
\ket{H \! \left[ \eta \right]} & =
	\begin{bmatrix}
	1 \\
	\log_2(3) - \sfrac{2}{3}  \\
	0 \\
	0
\end{bmatrix} .
\end{align*}

As a step in calculating $\EE$, $\SI$, and $\TI$ we find:
\begin{align*}
\sum_{\lambda \in \Lambda_W \atop |\lambda| < 1}
  \frac{1}{1-\lambda} \StartMS W_\lambda
& =
\begin{bmatrix}
  2 & \tfrac{3}{2} & - \tfrac{3}{2} & - 2
\end{bmatrix} , \text{ and } \\
\sum_{\lambda \in \Lambda_W \atop |\lambda| < 1}
  \frac{1}{(1-\lambda)^2} \StartMS W_\lambda
& =
\begin{bmatrix}
  6 & 6 & - 5 & - 7
\end{bmatrix} .
\end{align*}

Hence, for the scalar complexity measures of the strictly timelike subprocess, we find:
\begin{align*}
\hmu &= 1/2 ~\text{bit per step} , \\
\Cmu &= 1  ~\text{bit},  \\
\EE &= 1  ~\text{bit},  \\
\TI & = 1 + \tfrac{3}{2} \log_2(3) ~\text{bits-symbols}, \text{ and } \\
\SI & = 1 + \tfrac{3}{2} \log_2(3) ~\text{bits} .
\end{align*}

\subsection{Spacelike complexity}

\begin{figure}[h!]
  \centering
  \includegraphics[width=0.25\textwidth]{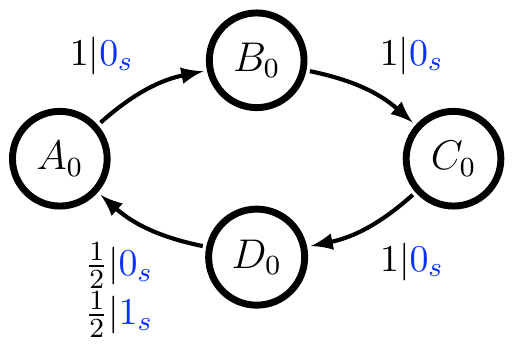}
\caption{\EM\ of one of the strictly spacelike subprocesses at $v = 1$.
  }
\label{fig:ECA22StrictlySpacelikeSubprocess}
\end{figure}

We just considered the complexity measures of the ECA 22 domain for one of the
strictly timelike subprocesses at $v = 0$. At the other velocity extreme of
$v = 1$, one of the strictly spacelike domain subprocesses is similar: the
noisy period-$4$ process shown in
Fig.\ \ref{fig:ECA22StrictlySpacelikeSubprocess}.

\begin{figure}[h!]
  \centering
  \includegraphics[width=0.45\textwidth]{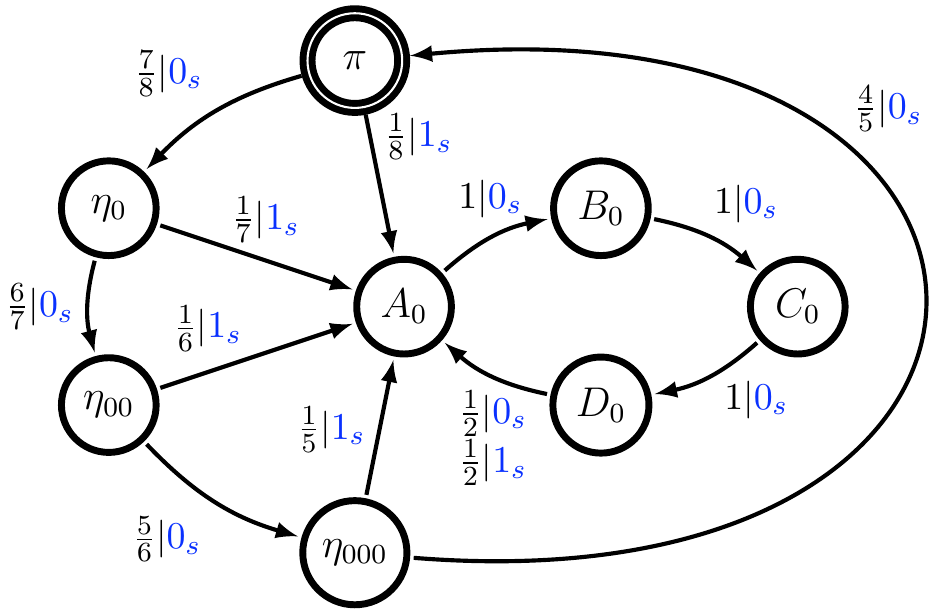}
\caption{MSP of the strictly spacelike subprocess of Fig.\ \ref{fig:ECA22StrictlySpacelikeSubprocess}.
  }
\label{fig:ECA22StrictlySpacelikeSubprocess_MSP}
\end{figure}

For this strictly spacelike subprocess, we obtain the MSP shown in Fig.\ \ref{fig:ECA22StrictlySpacelikeSubprocess_MSP},
where each mixed state is represented by a label corresponding to a particular
state distribution $\left[ \Pr(A_0), \Pr(B_0), \Pr(C_0), \Pr(D_0) \right]$.

The mixed states induced from the stationary distribution by observing
sequences can be cast into a column vector:
\begin{align}
\label{eq: ECA22 strictly spacelike mixed states}
\sum_{\eta \in \MxSSet} \eta \ket{\delta_\eta} & =
	\begin{bmatrix}
	 	\pi \\
		\eta_0 \\
		\eta_{00} \\
		\eta_{000} \\
		A_0 \\
		B_0 \\
		C_0 \\
		D_0
	\end{bmatrix}
	=
	\begin{bmatrix}
	\left( \sfrac{1}{4}, \right. 	& \sfrac{1}{4}, 		& \sfrac{1}{4},		& \left. \sfrac{1}{4} \right) \\
	\left( \sfrac{1}{7}, \right. 	& \sfrac{2}{7}, 		& \sfrac{2}{7},		& \left. \sfrac{2}{7} \right) \\
	\left( \sfrac{1}{6}, \right. 	& \sfrac{1}{6}, 		& \sfrac{1}{3},		& \left. \sfrac{1}{3} \right) \\
	\left( \sfrac{1}{5}, \right. 	& \sfrac{1}{5}, 		& \sfrac{1}{5},		& \left. \sfrac{2}{5} \right) \\
	\left( 1, \right. 			& 0, 				& 0,				& \left. 0 \right) \\
	\left( 0, \right. 			& 1, 				& 0,				& \left. 0 \right) \\
	\left( 0, \right. 			& 0, 				& 1,				& \left. 0 \right) \\
	\left( 0, \right. 			& 0, 				& 0,				& \left. 1 \right) \\
\end{bmatrix} .
\end{align}

Using the same ordering of mixed states as in Eq.\ \eqref{eq: ECA22 strictly spacelike mixed states},
the mixed-state presentation has state-transition matrix:
\begin{align*}
W = \sum_{x \in \mathcal{A}} W^{(x)}  =
\begin{bmatrix}
0 			& \sfrac{7}{8} 	& 0			& 0 			& \sfrac{1}{8}  	& 0	& 0	& 0 	\\
0			& 0 			&\sfrac{6}{7} 	& 0 			& \sfrac{1}{7} 	& 0	& 0 	& 0 	\\
0			& 0 			& 0 			& \sfrac{5}{6} 	& \sfrac{1}{6} 	& 0	& 0 	& 0	\\
\sfrac{4}{5}	& 0 			& 0  			& 0 			& \sfrac{1}{5} 	& 0	& 0 	& 0	\\
0 			& 0 			& 0  			& 0 			& 0 			& 1	& 0 	& 0 	\\
0 			& 0 			& 0  			& 0 			& 0 			& 0	& 1 	& 0 	\\
0 			& 0 			& 0  			& 0 			& 0 			& 0	& 0 	& 1 	\\
0 			& 0 			& 0  			& 0 			& 1 			& 0	& 0 	& 0 	\\
\end{bmatrix} .
\end{align*}

From Eq.\ \eqref{eq: ECA22 strictly spacelike mixed states}, we have
the internal entropy of each mixed state:
\begin{align*}
\ket{H \! \left[ \eta \right]} & =
	\begin{bmatrix}
	2 \\
	\log_2(7) - \sfrac{6}{3}  \\
	\log_2(3) + \sfrac{1}{3}  \\
	\log_2(5) - \sfrac{2}{5}  \\
	0 \\
	0 \\
	0 \\
	0
\end{bmatrix} .
\end{align*}

From Fig.\ \ref{fig:ECA22StrictlySpacelikeSubprocess_MSP}, we obtain the
entropy of transitioning from each mixed state:
\begin{align*}
\ket{H(W^\Abet)} \vphantom{\bigl[ \bigr]}
	=
	\begin{bmatrix}
	3 - \frac{7}{8} \log_2(7) \\
	 \log_2(7) - \frac{6}{7} \log_2(3) - \sfrac{6}{7} \\
	1 + \log_2(3) - \frac{5}{6} \log_2(5)  \\
	\log_2(5) - \sfrac{8}{5} \\
	0 \\
	0 \\
	0 \\
	1
	\end{bmatrix}
	~.
\end{align*}

Solving $\det (\lambda I - W) = 0$ gives $W$'s eigenvalues,
the fourth roots of unity and the fourth roots of $\tfrac{1}{2}$:
\begin{align*}
\Lambda_W &=
	\left\{
	1, \,
	-1, \,
	i, \,
	-i, \,
	\sqrt[4]{\tfrac{1}{2}}, \,
	- \sqrt[4]{\tfrac{1}{2}}, \,
	i \sqrt[4]{\tfrac{1}{2}},  \,
	- i \sqrt[4]{\tfrac{1}{2}}
  	\right\}
    ~.
\end{align*}

We obtain the projection operators $\{ W_\lambda \}$ as in
Eq.\ \eqref{eq:ProjOperatorsViaResiduesOfResolvent} via the residues of the
$W$'s resolvent around each eigenvalue in $\Lambda_W$, which are also the
isolated poles of $W$'s resolvent. In fact, we only obtain the first row
$\{ \StartMS  W_\lambda \}$ of each projection operator, since finding the
entire resolvent and set of projection matrices is superfluous to our immediate
goal.

Employing a complex variable $z$, matrix inversion gives the first row of the
resolvent matrix:
\begin{align}
\label{eq:Resolvent1stLine4SpacelikeECAexample}
\StartMS & (zI - W)^{-1} \\
  & = \frac{1}{8(z^4 - \sfrac{1}{2})}
    \begin{bmatrix}
	\frac{z^3}{8}
	& 7 z^2
	& 6 z
	& 5
	& \frac{z^3}{z-1}
	& \frac{z^2}{z-1}
	& \frac{z}{z-1}
	& \frac{1}{z-1}
    \end{bmatrix}
	\nonumber
	~.
\end{align}
The poles of each element of $\StartMS (zI - W)^{-1}$ are immediately evident:
The first four elements only have poles at the four fourth roots of
$\sfrac{1}{2}$. The last four entries have five poles---the four fourth roots
of $\sfrac{1}{2}$ and at $z$ equal to unity.

Since all of the poles are simple ($W$ being diagonalizable), the projection
operators can be most easily obtained by the residue algorithm for simple poles:
\begin{align*}
W_\lambda
& = \text{Res} \left(  (zI - W)^{-1} , \;  z \to\lambda  \right) \\
& =  \lim_{z \to \lambda}  \,  (z - \lambda)  (zI - W)^{-1} ,
\end{align*}
where the residues are taken element-wise. As a simple consequence:
\begin{align*}
\StartMS W_\lambda
& = \text{Res} \left(  \StartMS (zI - W)^{-1} , \;  z \to\lambda  \right) \\
& =  \lim_{z \to \lambda}  \,  (z - \lambda)  \StartMS (zI - W)^{-1} .
\end{align*}

Hence, we immediately see that the first row of all projection operators
associated with the three roots of unity, besides unity itself, are row
vectors of all zeros:
\begin{align*}
\StartMS W_{-1}
& = \begin{bmatrix} 0 & 0 & 0 & 0 & 0 & 0 & 0 & 0 \end{bmatrix} \\
\StartMS W_{i}
& = \begin{bmatrix} 0 & 0 & 0 & 0 & 0 & 0 & 0 & 0 \end{bmatrix} \\
\StartMS W_{-i}
& = \begin{bmatrix} 0 & 0 & 0 & 0 & 0 & 0 & 0 & 0 \end{bmatrix} .
\end{align*}
Moreover, the first row of the projection operator associated with unity, which is also identifiable with the stationary distribution over the mixed-state presentation, is easily found to be:
\begin{align*}
\StartMS W_{1}
& = \begin{bmatrix} 0 & 0 & 0 & 0 & \tfrac{1}{4} & \tfrac{1}{4} & \tfrac{1}{4} & \tfrac{1}{4} \end{bmatrix} .
\end{align*}

The remaining four projection operators are all associated with eigenvalues $\lambda$ such that $\lambda^4 = \sfrac{1}{2}$. To obtain the remaining
residues of Eq.\ \eqref{eq:Resolvent1stLine4SpacelikeECAexample}, we note
that:
\begin{align*}
\frac{z-\lambda}{z^4 - \lambda^4} = \frac{1}{z^3 + \lambda z^2 + \lambda^2 z + \lambda^3}
  ~,
\end{align*}
so that
\begin{align*}
\lim_{z \to \lambda} \left( \frac{z-\lambda}{z^4 - \lambda^4} \right) = \frac{1}{4} \lambda^{-3} .
\end{align*}
The first row of the remaining four projection operators is thus:
\begin{align*}
\StartMS & W_{\lambda} = \\
& \frac{1}{8}
\begin{bmatrix}
	2 &
	\frac{7}{4 \lambda} &
	\frac{3}{2 \lambda^2} &
	\frac{5}{4 \lambda^3} &
	\frac{-1}{1-\lambda} &
	\frac{-1}{\lambda (1-\lambda)} &
	\frac{-1}{\lambda^2 (1-\lambda)} &
	\frac{-1}{\lambda^2 (1-\lambda)}
    \end{bmatrix}
  ~,
\end{align*}
for $\lambda^4 = \sfrac{1}{2}$.

It is then straightforward to calculate the complexity measures:
\begin{align*}
\hmu &= 1/4 ~\text{bit per step} , \\
\Cmu &= 2  ~\text{bits},  \\
\EE &= 2  ~\text{bit},  \\
\TI   & = \tfrac{5}{2} + \tfrac{7}{4} \log_2(7) + \tfrac{5}{4} \log_2(5) + \tfrac{3}{2} \log_2(3) ~\text{bits-symbols} , \\
\text{and} & \\
\SI   & = \tfrac{5}{2} + \tfrac{7}{4} \log_2(7) + \tfrac{5}{4} \log_2(5) + \tfrac{3}{2} \log_2(3) ~\text{bits} .
\end{align*}

Since $\EE = \Cmu$, each the timelike and spacelike subprocess shares all
information that is stored from their \emph{past} with their \emph{future}
via their \emph{present}. In other words, the subprocesses have no crypticity:
$\PC = \Cmu - \EE = 0$.

We conclude that the strictly spacelike subprocess is less random (via
$\hmu$), stores more information (via $\Cmu$), shares more information
with the future (via $\EE$), and is more difficult to synchronize to
(via $\SI$ and $\TI$) than the strictly timelike subprocess. Different
facets of complexity express themselves more or less prominently along
different spacetime paths within the same spacetime domain.

%

\section{Zinc Sulfide: Intrinsic Spatial Computation in Polytypic Materials}

\begin{figure}[h!]
  \centering
  \includegraphics[width=\columnwidth]{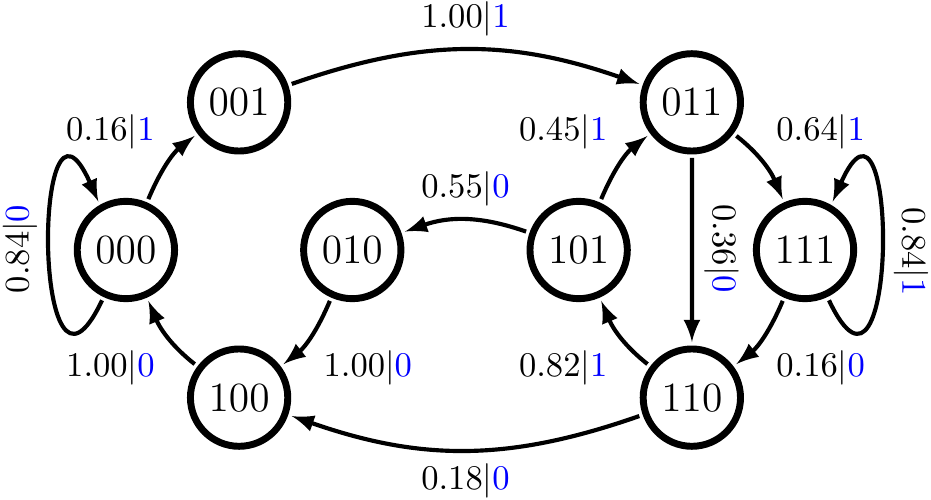}
\caption{\EM\ $\mathcal{M}$ that captures the spatial organization in
  sample SK135 of Zinc Sulfide, expressed in the Hagg notation.
  Repeated here from Ref. [A3] with permission.
  }
\label{fig:ZnSeM}
\end{figure}

As a final example we analyze the intrinsic computation in the spatial
organization of the polytypic, closed-packed material Zinc Sulfide.
Using experimentally measure X-ray diffraction spectra, Ref. [A3]
extracted the \eM\ shown in Fig. \ref{fig:ZnSeM}.
This is for the sample called SK135 there.

$\mathcal{M}$ has the two-symbol alphabet $\mathcal{A} = \{ 0, 1 \}$ with the
corresponding symbol-labeled transition matrices:
\begin{align*}
T^{(0)} & =
    \begin{bmatrix}
    0.84 & \phantom{0}0\phantom{0} & 0 & \phantom{0}0\phantom{0} & 0 & \phantom{0}0\phantom{0} & 0 & \phantom{0}0\phantom{0} \\
    0 & 0 & 0 & 0 & 0 & 0 & 0 & 0 \\
    0 & 0 & 0 & 0 & 1 & 0 & 0 & 0 \\
    0 & 0 & 0 & 0 & 0 & 0 & 0.36 & 0 \\
    1 & 0 & 0 & 0 & 0 & 0 & 0 & 0 \\
    0 & 0 & 0.55 & 0 & 0 & 0 & 0 & 0 \\
    0 & 0 & 0 & 0 & 0.18 & 0 & 0 & 0 \\
    0 & 0 & 0 & 0 & 0 & 0 & 0.16 & 0
    \end{bmatrix}
\text{and } \\
T^{(1)} & =
    \begin{bmatrix}
    \phantom{0}0\phantom{0} & 0.16 & \phantom{0}0\phantom{0} & 0 & \phantom{0}0\phantom{0} & 0 & \phantom{0}0\phantom{0} & 0 \\
    0 & 0 & 0 & 1 & 0 & 0 & 0 & 0 \\
    0 & 0 & 0 & 0 & 0 & 0 & 0 & 0 \\
    0 & 0 & 0 & 0 & 0 & 0 & 0 & 0.64 \\
    0 & 0 & 0 & 0 & 0 & 0 & 0 & 0 \\
    0 & 0 & 0 & 0.45 & 0 & 0 & 0 & 0 \\
    0 & 0 & 0 & 0 & 0 & 0.82 & 0 & 0 \\
    0 & 0 & 0 & 0 & 0 & 0 & 0 & 0.84
    \end{bmatrix}
  ~.
\end{align*}
We find the stationary distribution from the state-transition matrix $T$:
\begin{align*}
\bra{\pi} \approx \begin{bmatrix} 0.32 & 0.05 & 0.04 & 0.08 & 0.05 & 0.07 & 0.08 & 0.32 \end{bmatrix} ~.
\end{align*}

The mixed-state presentation $\mathcal{M}_\text{msp}$ gives the dynamics
induced by observed symbols over $\mathcal{M}$'s state distributions,
starting from the stationary distribution $\pi$.

\begin{figure}[h!]
  \centering
  \includegraphics[width=0.48\textwidth]{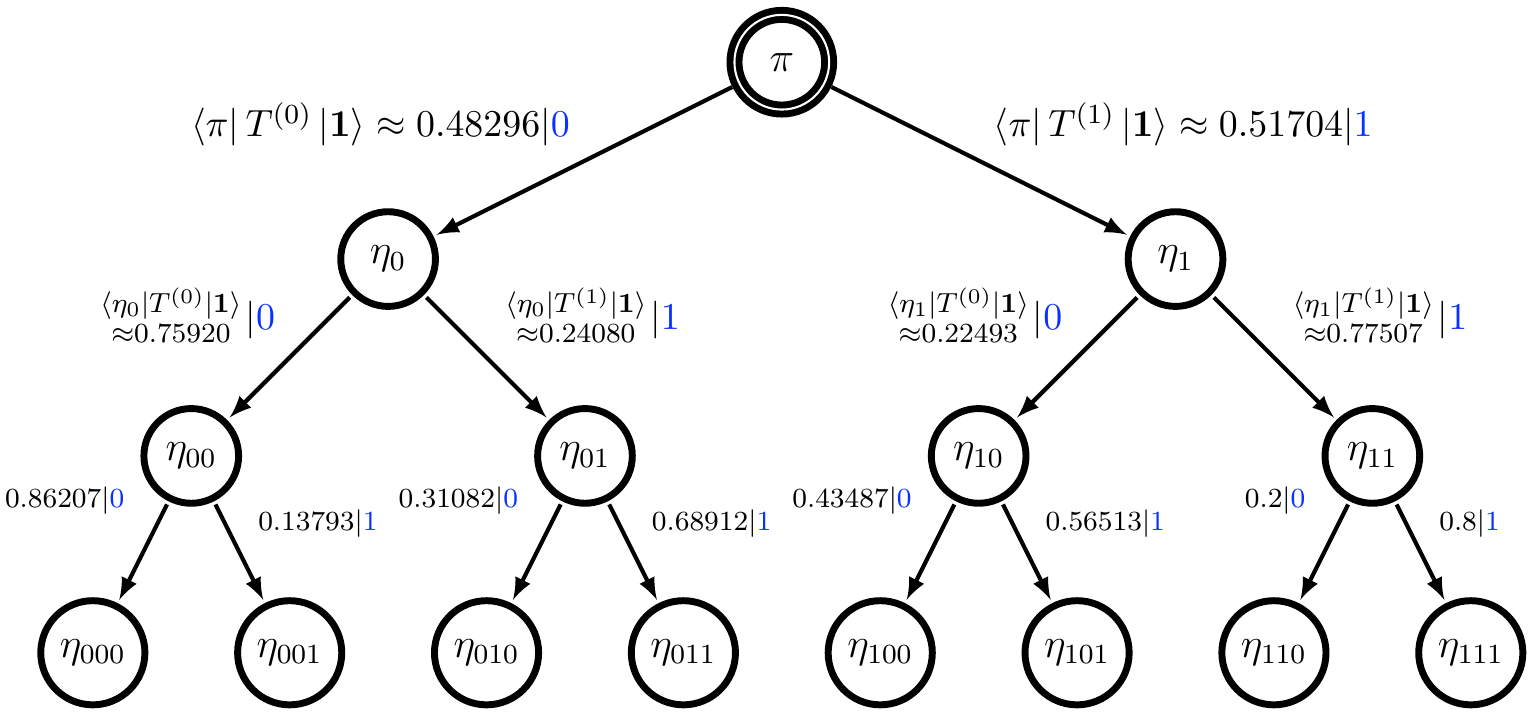}
\caption{The tree-like transition structure among the transient mixed
  states. The transition structure among the recurrent states (those states in
  the bottom row) is isomorphic to the recurrent structure of
  Fig. \ref{fig:ZnSeM}.
  }
\label{fig:ZnSMSP_TransientTree}
\end{figure}

Since $\mathcal{M}$ has finite Markov order of $R = 3$, the mixed-state
presentation is a depth-$3$ tree of nonrecurrent transient states that feed
into the recurrent states of $\mathcal{M}$, as shown in Fig.
\ref{fig:ZnSMSP_TransientTree}. Algebraically, this translates to the addition
of a zero eigenvalue with index $\nu_0 = 3$. Hence, $W$ is nondiagonalizable,
although $a_\lambda = g_\lambda$ for all $\lambda \neq 0$.
Eq.\ \eqref{eq:GeneralSpectralDecomp} then implies that:
\begin{align}
\label{eq: almost diagonalizable W^L spectral decomp}
W^L
& =
\Biggl\{
    \sum_{\lambda \in \Lambda_W}
    \lambda^L  W_\lambda
\Biggr\}
+
\sum_{N=1}^{\nu_0 - 1}  \delta_{L, N}  W_0 W^N
.
\end{align}
Moreover, the complexity measures involve $\StartMS  W^L $ and thus $\StartMS  W_{\lambda} $.
Importantly, in this example,
$\StartMS W_{\lambda} = \bra{\mathbf{0}}$ for all $0 < | \lambda | < 1$.
In particular,
$\StartMS W_{0} = \StartMS (I - W_1)$ and
$\StartMS W_1 = \StartMS  \one \rangle \bra{\pi_W} = \bra{\pi_W} $.
This is shown explicitly in Table \ref{tab:ZnSMuWEtc}~.

\begin{table}
  \centering
\begin{tabular}{crc}
  \hline
$\lambda$  &    $\; \nu_\lambda$  &  $\qquad \qquad
\StartMS W_{\lambda} $  \\
  \hline \\
$1$ & $1$
    & $ \quad \begin{matrix} \bigl[ \bigr. \phantom{-}0 &
    0 & 0 & 0 & 0 \\
    \ldots 0 & 0 & \phantom{-}0.32 & \phantom{-}0.05 & \phantom{-}0.04 \\
    \phantom{-}\ldots 0.08 & \phantom{-}0.05 & \phantom{-}0.07 & \phantom{-}0.08 & \phantom{-}0.32 \bigl. \bigr] \end{matrix} $     \\ \\
 $0$ & $3$
    & $ \quad \begin{matrix} \bigl[ \bigr. \phantom{- -}1 &
    0 & 0 & 0 & 0 \\
    \ldots \phantom{-}0 & 0 & -0.32 & -0.05 & -0.04 \\
    \ldots -0.08 & -0.05 & -0.07 & -0.08 & -0.32 \bigl. \bigr] \end{matrix} $       \\ \\
$\zeta \in \Lambda_W^{\setminus \{0, 1\}}$
& $1$ & $\bra{ \mathbf{0} }$ \\
\hline
\end{tabular}
\caption{
  Useful spectral quantities for the ZnS polytype analysis.
  For compactness we define
  $\Lambda_W^{\setminus \{0, 1\}} \equiv \Lambda_W \setminus \{0, 1\}$ to be
  the set of $W$'s eigenvalues that are not zero or unity. None of the
  projection operators associated with these other eigenvalues overlap with
  $\StartMS$. Moreover, note that $\StartMS W_0 = \StartMS (I - W_1)$ and
  $\StartMS W_1 = \StartMS  \one \rangle \bra{\pi_W} = \bra{\pi_W}$.
  }
\label{tab:ZnSMuWEtc}
\end{table}

Different complexity measures track the evolution of different types of information in (or about) the system.
The entropy of transitioning from the various states of uncertainty is given by the ket $\ket{H(W^\Abet)}$, whereas the internal entropy of the states of uncertainty themselves is given by the ket $\ket{H \! \left[ \eta \right]}$.
From the labeled transition matrices of the mixed state presentation, we find:
\begin{align*}
\ket{H(W^\Abet)} & \approx
    \begin{bmatrix}
    {\color{blue} H_2(0.48296) } \\
    {\color{blue} H_2(0.24080) } \\
    {\color{blue} H_2(0.22493) } \\
    {\color{blue} H_2(0.13793) } \\
    {\color{blue} H_2(0.31082) } \\
    {\color{blue} H_2(0.43487) } \\
    {\color{blue} H_2(0.20000) } \\
    H_2(0.16)  \\
    0 \\
    0  \\
    H_2(0.36) \\
    0  \\
    H_2(0.45) \\
    H_2(0.18)  \\
    H_2(0.16)
    \end{bmatrix}
    \; \approx \;
    \begin{bmatrix}
    {\color{blue} 0.9992 } \\
    {\color{blue} 0.7964 } \\
    {\color{blue} 0.7691 } \\
    {\color{blue} 0.5788 } \\
    {\color{blue} 0.8941 } \\
    {\color{blue} 0.9877 } \\
    {\color{blue} 0.7219 } \\
    0.6343  \\
    0 \\
    0  \\
    0.9427 \\
    0  \\
    0.9928 \\
    0.6801  \\
    0.6343
    \end{bmatrix}  ,
\end{align*}
where $H_2(q)$ is the binary entropy function, $H_2(q) \equiv - q \log_2(q) - (1 - q) \log_2(1-q)$, and quantities associated with the transient states are colored blue.  And from the mixed states themselves,
\begin{align*}
\sum_{\eta \in \MxSSet} & \eta \ket{\delta_\eta} \\
& \approx
    \begin{bmatrix}
{\color{blue}   \left( 0.32 \right. } &     {\color{blue} 0.05 } &  {\color{blue} 0.04 } &  {\color{blue} 0.08 } &  {\color{blue} 0.05 } &  {\color{blue} 0.07 } &  {\color{blue} 0.08} &   {\color{blue} 0.32   \left. \right) } \\
{\color{blue}   \left( \right. 0.65 } &     {\color{blue} 0 } &         {\color{blue} 0.07 } &  {\color{blue} 0 } &         {\color{blue} 0.1 } &       {\color{blue} 0 } &         {\color{blue} 0.17 } &  {\color{blue} 0      \left. \right) } \\
{\color{blue}   \left( \right.  0 } &       {\color{blue} 0.1 } &       {\color{blue} 0 } &     {\color{blue} 0.16 } &  {\color{blue} 0 } &     {\color{blue} 0.13 } &  {\color{blue} 0 } &         {\color{blue} 0.62 \left. \right) } \\
{\color{blue}   \left( \right. 0.86 } &     {\color{blue} 0 } &         {\color{blue} 0 } &     {\color{blue} 0 } &         {\color{blue} 0.14 } &  {\color{blue} 0 } &         {\color{blue} 0 } &         {\color{blue} 0  \left. \right) } \\
{\color{blue}   \left( \right. 0 } &        {\color{blue} 0.43 } &  {\color{blue} 0 } &     {\color{blue} 0 } &         {\color{blue} 0 } &     {\color{blue} 0.57 } &  {\color{blue} 0} &      {\color{blue} 0 \left. \right) } \\
{\color{blue}   \left( \right. 0 } &        {\color{blue} 0 } &         {\color{blue} 0.31 } &  {\color{blue} 0 } &         {\color{blue} 0 } &     {\color{blue} 0 } &         {\color{blue} 0.69 } &  {\color{blue} 0  \left. \right)} \\
{\color{blue}   \left( \right. 0 } &        {\color{blue} 0 } &         {\color{blue} 0 } &     {\color{blue} 0.2 } &       {\color{blue} 0 } &     {\color{blue} 0 } &         {\color{blue} 0 } &         {\color{blue} 0.8  \left. \right) } \\
    ( 1&        0&      0&      0&      0&      0&      0&      0  ) \\
    ( 0&        1&      0&      0&      0&      0&      0&      0  ) \\
    ( 0&        0&      1&      0&      0&      0&      0&      0  ) \\
    ( 0&        0&      0&      1&      0&      0&      0&      0  ) \\
    ( 0&        0&      0&      0&      1&      0&      0&      0  ) \\
    ( 0&        0&      0&      0&      0&      1&      0&      0  ) \\
    ( 0&        0&      0&      0&      0&      0&      1&      0  ) \\
    ( 0&        0&      0&      0&      0&      0&      0&      1  )
\end{bmatrix} ,
\end{align*}
we have:
\begin{align*}
\ket{H \! \left[ \eta \right]} & \approx
    \begin{bmatrix}
    {\color{blue} 2.5018 } \\
    {\color{blue} 1.4511 } \\
    {\color{blue} 1.5508 } \\
    {\color{blue} 0.5788 } \\
    {\color{blue} 0.9877 } \\
    {\color{blue} 0.8941 } \\
    {\color{blue} 0.7219 } \\
    0  \\
    0 \\
    0  \\
    0 \\
    0  \\
    0 \\
    0  \\
    0
    \end{bmatrix}   .
\end{align*}

From the above, we have
for the finite-$L$ entropy rate convergence:
\begin{align*}
h_\mu(L)
& =
\sum_{N=1}^{\nu_0 - 1}  \delta_{L-1, N}  \StartMS W_0 W^N | H(W^\Abet) \rangle \nonumber \\
& \quad +
\sum_{\lambda \in \Lambda_W} \lambda^{L-1}
  \StartMS  W_{\lambda}  | H(W^\Abet) \rangle \\
& =
\sum_{N=0}^{2}  \delta_{L-1, N} \overbrace{\StartMS W_0}^{= \StartMS - \bra{\pi_W}} W^N | H(W^\Abet) \rangle \nonumber \\
& \quad +
\underbrace{\StartMS  W_1}_{= \bra{\pi_W}}  | H(W^\Abet) \rangle \\
& =
\hmu +
\sum_{N=0}^{2}  \delta_{L-1, N} \left( \StartMS W^N | H(W^\Abet) \rangle - \hmu \right)  \\
& =
\delta_{L, 1}   \StartMS H(W^\Abet) \rangle
+
\delta_{L, 2}   \StartMS W | H(W^\Abet) \rangle \nonumber \\
& \qquad +
\delta_{L, 3}   \StartMS W^2 | H(W^\Abet) \rangle
+
u_{L-4} \hmu \\
\nonumber \\
& \approx
0.999 \delta_{L, 1}
+
0.782 \delta_{L, 2}
+
0.720 \delta_{L, 3}
+
0.599 u_{L-4} ~,
\end{align*}
where $u_{L-4}$ is the unit step sequence that is zero for $L < 4$ and unity for $L \geq 4$.

For the excess entropy, we find:
\begin{align*}
\EE
& =
\sum_{N=1}^{\nu_0 - 1}   \StartMS W_0 W^N | H(W^\Abet) \rangle \nonumber \\
& \quad +
\sum_{\lambda \in \Lambda_W \atop | \lambda | < 1} \frac{1}{1 - \lambda}
  \StartMS  W_{\lambda}  | H(W^\Abet) \rangle \\
& =
\sum_{N=0}^{2}  \overbrace{\StartMS W_0}^{= \StartMS - \bra{\pi_W}} W^N | H(W^\Abet) \rangle \\
& =
- 3 \bra{\pi_W} H(W^\Abet) \rangle +   \sum_{N=0}^{2}  \StartMS W^N | H(W^\Abet) \rangle \\
& = \StartMS (I + W + W^2) | H(W^\Abet) \rangle - 3 \hmu \\
\nonumber \\
& \approx 0.70430 ~.
\end{align*}

For the transient information, we have:
\begin{align*}
\TI
& =
\sum_{N=1}^{\nu_0 - 1}  (N+1) \StartMS W_0 W^N | H(W^\Abet) \rangle \nonumber \\
& \quad +
\sum_{\lambda \in \Lambda_W \atop | \lambda | < 1} \frac{1}{(1 - \lambda)^2}
  \StartMS  W_{\lambda}  | H(W^\Abet) \rangle \\
& =
\sum_{N=0}^{2}  \overbrace{\StartMS W_0}^{= \StartMS - \bra{\pi_W}} (N+1) W^N | H(W^\Abet) \rangle \\
& =
- 6 \bra{\pi_W} H(W^\Abet) \rangle +   \sum_{N=0}^{2}  \StartMS (N+1) W^N | H(W^\Abet) \rangle \\
& = \StartMS (I + 2 W + 3 W^2) | H(W^\Abet) \rangle - 6 \hmu \\
\nonumber \\
& \approx 1.12982 ~.
\end{align*}

The synchronization information is:
\begin{align*}
\SI
& =
\sum_{N=1}^{\nu_0 - 1}   \StartMS W_0 W^N | H[\eta] \rangle \nonumber \\
& \quad +
\sum_{\lambda \in \Lambda_W \atop | \lambda | < 1} \frac{1}{1 - \lambda}
  \StartMS  W_{\lambda}  | H[\eta] \rangle \\
& =
\sum_{N=0}^{2}  \overbrace{\StartMS W_0}^{= \StartMS - \bra{\pi_W}} W^N | H[\eta] \rangle \\
& =
- 3 \bra{\pi_W} H[\eta] \rangle +   \sum_{N=0}^{2}  \StartMS W^N | H[\eta] \rangle \\
& = \StartMS (I + W + W^2) | H[\eta] \rangle  \\
\nonumber \\
& \approx 4.72481 ~.
\end{align*}

Collecting our results, the scalar complexity measures are:
\begin{align*}
\hmu & = 0.59916 ~\text{bits per step} , \\
\Cmu & = 2.50179 ~\text{bits},  \\
\EE  & = 0.70430 ~\text{bits},  \\
\TI  & = 1.12982 ~\text{bits-symbols}, \text{ and } \\
\SI  & = 4.72481 ~\text{bits} .
\end{align*}

\section*{References}

[A1] R.~G. James, K.~Burke, and J.~P. Crutchfield.
\newblock Chaos forgets and remembers: {Measuring} information creation,
  destruction, and storage.
\newblock in preparation.

[A2] J.~E. Hanson and J.~P. Crutchfield.
\newblock {\em Physica D}, 103:169--189, 1997.

[A3] D.~P. Varn, G.~S. Canright, and J.~P. Crutchfield.
\newblock {\em Phys. Rev. B}, 66(17):174110--3, 2002.

\end{document}